\documentclass[journal=jacsat,manuscript=article]{achemso}

\usepackage{chemformula} % Formula subscripts using \ch{}
\usepackage[T1]{fontenc} % Use modern font encodings
\usepackage{lmodern}
\usepackage[utf8]{inputenc}
\usepackage{amsmath}

\author{Lottie L. Murray}
\affiliation[UDMSEG]{Department of Materials Science and Engineering, University of Delaware, Newark, DE, 19716, US}
\author{Eric Herrmann}
\affiliation[UDMSEG]{Department of Materials Science and Engineering, University of Delaware, Newark, DE, 19716, US}
\author{Igor Evangelista}
\affiliation[UDMSEG]{Department of Materials Science and Engineering, University of Delaware, Newark, DE, 19716, US}
\author{Anderson Janotti}
\affiliation[UDMSEG]{Department of Materials Science and Engineering, University of Delaware, Newark, DE, 19716, US}
\author{Xi Wang}
\affiliation[UDMSEG]{Department of Materials Science and Engineering, University of Delaware, Newark, DE, 19716, US}
\author{Matthew F. Doty}
\affiliation[UDMSEG]{Department of Materials Science and Engineering, University of Delaware, Newark, DE, 19716, US}
\email{doty@udel.edu}

\title{Quantifying the Relationship Between Strain and Bandgap in Thin Ga$_2$Se$_2$}

\begin{document}

%\begin{tocentry}

%Some journals require a graphical entry for the Table of Contents.
%This should be laid out ``print ready'' so that the sizing of the

%Inside the \texttt{tocentry} environment, the font used is Helvetica
%8\,pt, as required by \emph{Journal of the American Chemical

%The surrounding frame is 9\,cm by 3.5\,cm, which is the maximum
%permitted for  \emph{Journal of the American Chemical Society}
%graphical table of content entries. The box will not resize if the
%content is too big: instead it will overflow the edge of the box.

%This box and the associated title will always be printed on a
%separate page at the end of the document.

%\end{tocentry}

\begin{abstract}
  We present a rigorous analysis that combines theory, simulation, and experimental measurements to quantify the relationship between strain and bandgap in two dimensional gallium selenide (Ga$_2$Se$_2$). Experimentally, we transfer thin Ga$_2$Se$_2$ flakes onto patterned substrates to deterministically induce multiaxial localized strain. We quantify the local strain using a combination of atomic force microscopy (AFM) measurements and COMSOL Multiphysics simulation. We then experimentally map the strain-induced bandgap shifts using high-resolution hyperspectral PL imaging to generate a robust and statistically significant dataset. We systematically fit this data to extract gauge factors that relate the bandgap shift to the local uniaxial and biaxial strain. We then compute the uniaxial and biaxial strain gauge factors via density functional theory (DFT) and find excellent agreement with the experimentally-determined values. Finally, we show that a simple model that computes bandgap shifts from the local uniaxial and biaxial strain predicts the observed multiaxial bandgap shift with less than 10\% error. The combined results provide a framework for deterministic realization of tailored bandgap profiles induced by controlled strain applied to Ga$_2$Se$_2$, with implications for the future realization of localized quantum emitters for quantum photonic applications.
\end{abstract}

\vspace{1em}


Two dimensional (2D) materials are of interest for both classical and quantum photonic applications because of their unique and tunable optical and electronic properties. For example, the 2D confinement of excitons leads to high exciton binding energies and enhanced nonlinear interactions\cite{excitons,exciton02,2dspes} with optical and electronic properties that can be tuned by both strain and applied electric fields \cite{strain_review,strain_gase,electric}. These properties can be exploited for classical optoelectronic devices such as modulators, photodetectors, lasers, and LEDs \cite{applications,applications02}. For quantum photonic applications, one important need is atomic-like quantum emitters that can provide a deterministic source of single photons or host a matter-based qubit (e.g.~electron spin) that can be manipulated via light. Quantum emitters in the solid-state are of great interest for quantum networking \cite{network,network02}, secure communication via Quantum Key Distribution (QKD) \cite{qkd,qkd02,2dspes}, or quantum sensing \cite{sensing,sensing02}. Such quantum emitters can be created in two dimensional materials when applied strain reduces the bandgap and creates localizing potentials analogous to quantum dots. Deterministically creating these localized potentials in a 2D material via the strain induced by a patterned underlying substrate provides one path toward scalable production of solid-state quantum emitters \cite{induced_strain,induced_strain02,band_engineering}.

Gallium selenide (Ga$_2$Se$_2$) is a 2D material that has received relatively little attention in comparison to transition metal dichaclogenides (TMDCs) despite the fact that it shares useful properties of TMDCs while also offering distinct advantages. For example, Ga$_2$Se$_2$ transitions from an indirect to direct bandgap when containing more than approximately 7 tetralayers. Because monolayers are not required to achieve a direct bandgap, it is easier to create devices via exfoliation and transfer from Ga$_2$Se$_2$ \cite{ptmd, gase_structure}. Moreover, like TMDCs, strain can be used to modify the bandgap of multilayer Ga$_2$Se$_2$, which provides a unique platform for strain engineering, spatially-localized control over direct vs indirect bandgaps, and the creation of localized confining potentials \cite{band_engineering}. 
%As with TMDCs, strain modifies the bandgap of thin Ga$_2$Se$_2$ by changing the spacing between atoms, which provides a tool for 

Prior investigations of the relationship between strain and strain-induced bandgap modification as measured by photoluminescence (PL) were conducted by transferring exfoliated flakes of Ga$_2$Se$_2$ onto flexible substrates and bending the material to create wrinkles \cite{flex_strain}. The strain profile in such experiments was approximated using the radius of curvature of the wrinkle, and the bandgap shift was quantified by successive PL measurements parallel and perpendicular to the strain axis. While this method provided convincing initial evidence that the bandgap of Ga$_2$Se$_2$ depends on induced strain, the results were limited by three key factors. First, a wrinkle is only a uniaxial strain vector, which has limited impact on the local band structure \cite{strain_2D,strain_2D02}. In contrast, biaxial strain compresses and stretches the crystal lattice along both in-plane directions, which results in more complex shifts to the band structure \cite{strain_2D}. Second, the wrinkle approach is neither scalable nor deterministic for applying local strain. Third, prior studies used small data sets that limited the conclusions that could be drawn. Our approach overcomes all three of these limitations. Specifically, our control over substrate features allow us to intentionally create and study regions with induced uniaxial, biaxial, and multiaxial strain. We compute the full strain tensor at each spatial location on the sample and correlate the strain with the observed PL shift, generating a large data set (84,525 spectra) from which statistically-significant conclusions can be drawn. Finally, we find excellent agreement between the experimental conclusions and density functional theory (DFT) calculations. While we report results that quantify the relationship between bandgap and strain only for Ga$_2$Se$_2$, the methods developed and reported here can be applied to any optically-active 2D material.

%\begin{figure}
%  \includegraphics[width=\linewidth]{Figure1.pdf}
%  \caption{\textbf{caption}}
%  \label{fig:Fig1}
%\end{figure}

%\subsection{Summary of Approach}
%General overview of the process \\
%mention distant structures: rings and ridges
\section{Results and Discussion}

\begin{figure}[h]
  \includegraphics[width=0.9\linewidth]{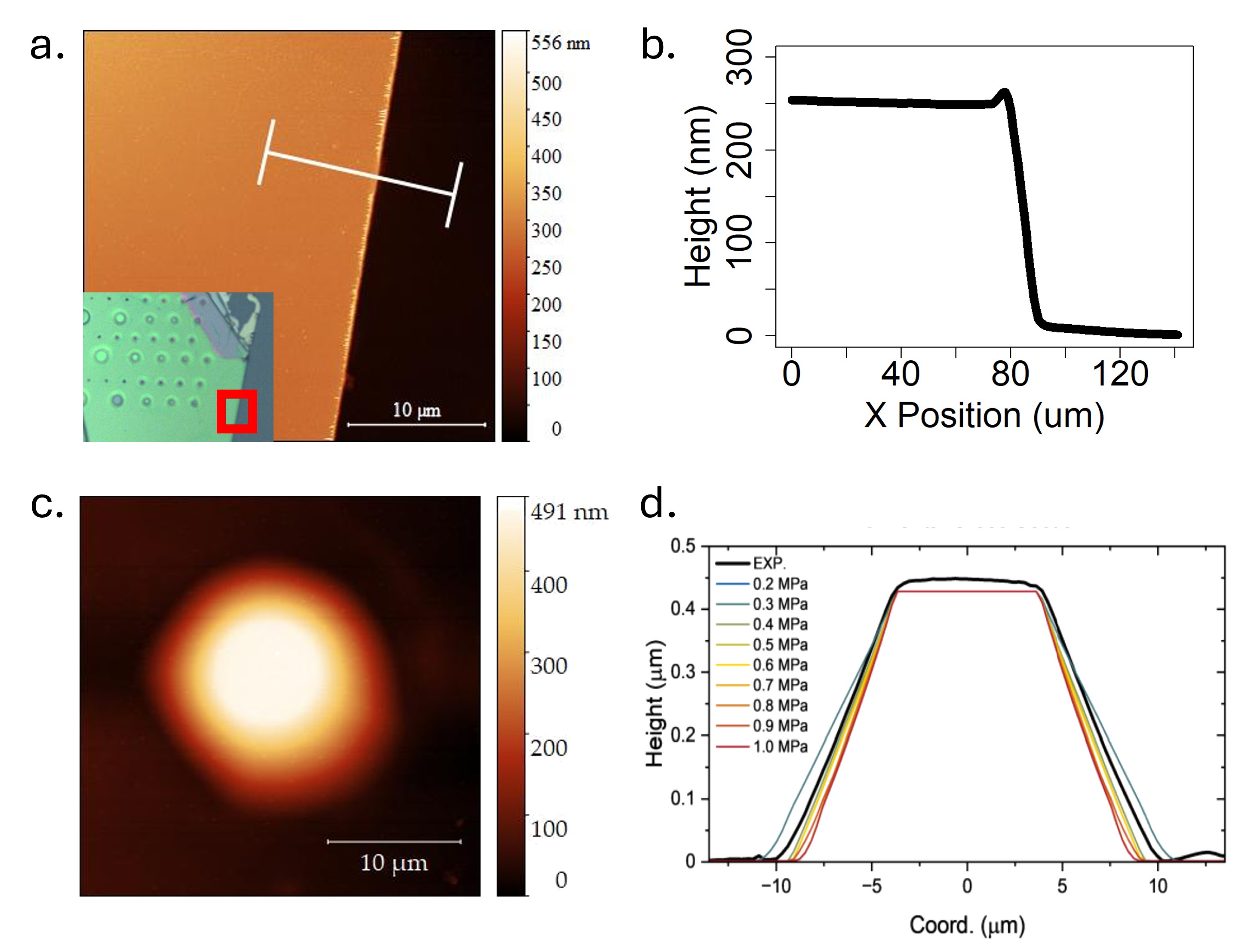}
  \caption{(a) AFM scan across the edge of a transferred flake, (b) Line profile of the AFM scan in (a) from which the Ga$_2$Se$_2$ thickness is extracted. (c) AFM scan of the membrane suspended over a ring structure. (d) Cross sectional line scans from the AFM data (black) and COMSOL simulations using various experimental loads (colors).}
  \label{afm_strain}
\end{figure}

We fabricated patterned substrates for applying localized strain by etching predefined structures on a silicon-on-insulator (SOI) substrate \cite{eric_fab}. To differentiate between the effects of biaxial and uniaxial strain, we fabricate two different types of structures: rings and parallel ridges. To isolate the effects of biaxial strain we tested 16 distinct ring structures, shown in Figure \ref{afm_strain}(a), with diameters ranging from 4$\mu$m to 12$\mu$m. To isolate the effects of uniaxial strain we tested 7 distinct ridge structures, shown in Figure \ref{afm_strain}(b), with ridge separation distances varying from 6$\mu$m to 14$\mu$m. Both rings and ridges protruded from the surface by about 430nm. Full details on the fabrication of these structures can be found in the SI.

\begin{figure}[h]
  \includegraphics[width=1\linewidth]{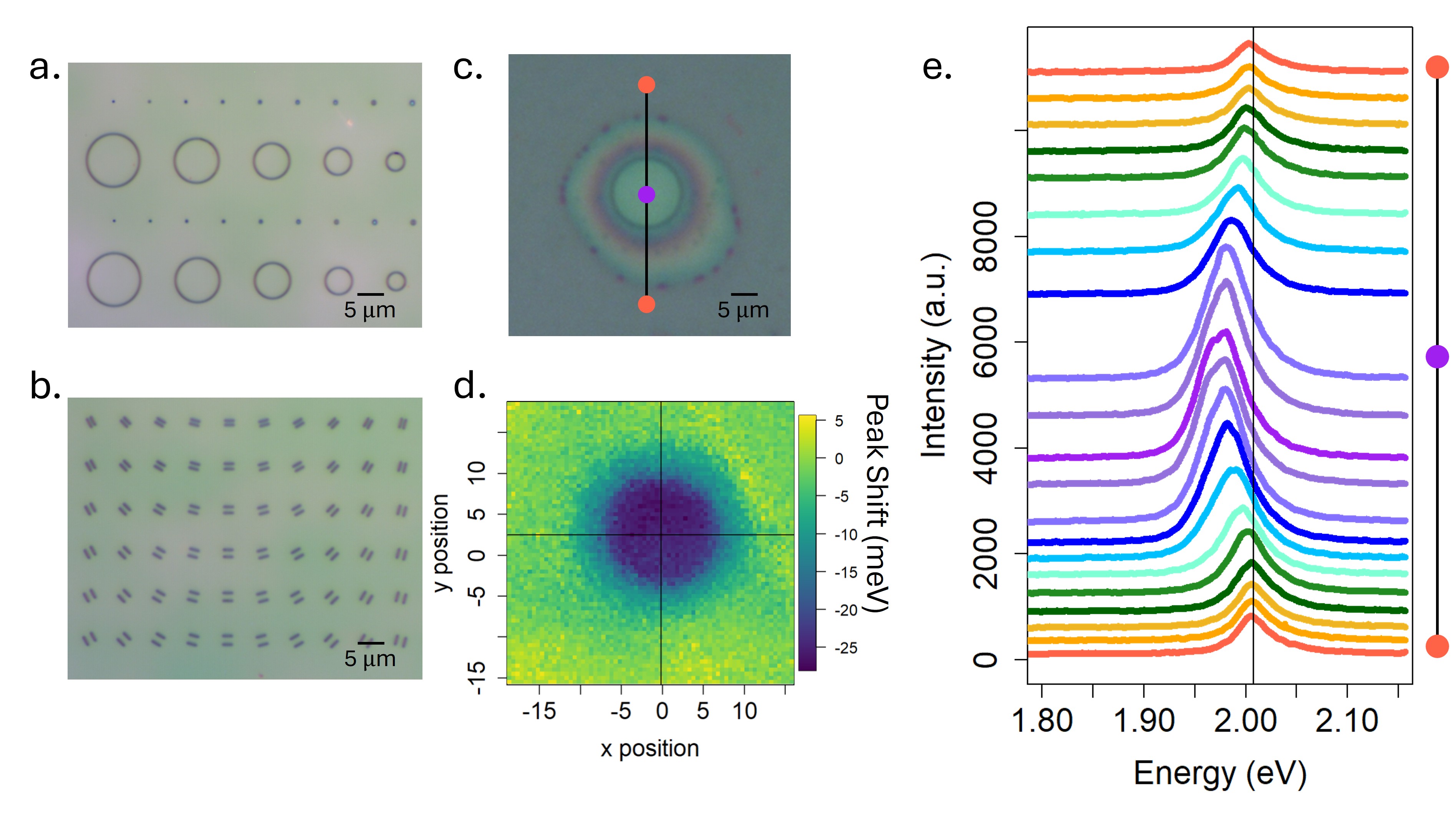}
  \caption{(a,b) Optical image of bare nanopatterned substrate with ring structures (a) and ridge structures (b). Optical image (c) and measured peak shift (d) for 2D Ga$_2$Se$_2$ suspended over a ring of $\sim$11 $\mu$m diameter. (e) Measured PL spectra as a function of position along the line indicated in (c).}
  \label{opt_ps}
\end{figure}

Thin flakes of Ga$_2$Se$_2$ are deterministically exfoliated and transferred onto the patterned substrates (Fig.~\ref{opt_ps}(a, b)), which induces strain in the suspended membranes. We conduct PL measurements at room temperature (298K) to obtain approximately 5000 measurements of the PL emission energy and intensity as a function of position around each feature, as illustrated in Fig.~\ref{opt_ps}d for a ring of 11$\mu$m diameter. In total we collect and analyze 84,525 spatially-resolved PL spectra. The shift in PL emission energy and the change in PL intensity as a function of position are evident in Fig.~\ref{opt_ps}e, which presents a subset of the PL spectra as function of position along the line indicated in Fig.~\ref{opt_ps}c. We next conduct atomic force microscopy (AFM) measurements on each structure to obtain a precise measure of the membrane thickness (Fig.~\ref{afm_strain}(a, b)) and a spatially-resolved profile of the deformation of the suspended membrane (Fig.~\ref{afm_strain}(c, d)). We compute the strain tensor as a function of position around each feature (Fig.~\ref{afm_strain}(e, f)) using COMSOL Multiphysics. We then correlate the spatially-resolved PL measurements with the strain tensor at each point to quantify the relationship between strain and bandgap shift. Finally, we conduct density functional theory (DFT) calculations of both the shifts in bandgap and the changes in PL intensity that are expected as a function of strain induced in Ga$_2$Se$_2$ and compare the DFT results to the experimental data.

%\subsection{Strain vs PL Shift}
%Combine strain and PL data\\
%results\\
%discussion\\
%- uniaxial vs biaxial strain
%- quantifying expt. PL shift vs strain eqn.

A critical element of our approach is that the parameters used in the COMSOL calculation of the spatially-resolved strain tensor are uniquely determined by requiring the computed deformation to match the spatial deformation of the membrane measured experimentally by AFM. As described in detail in the SI, the simulation method applies a boundary load that brings the 2D membrane into contact with both the ring/ridge features and the underlying flat substrate. The only free parameter in the simulation is the magnitude of this boundary load. The appropriate boundary load for each individual feature is determined by minimizing the root-mean-square deviation between the computed and the measured deformation profile. For example, the black line in Fig.~\ref{afm_strain}d shows a cross section of the experimentally-measured deformation of the membrane suspended over a ring structure. The colored lines show COMSOL simulations for a range of boundary load values; the best fit is obtained for a boundary load of 0.4 MPa. 

Once an optimized boundary load is determined, the strain tensor as a function of position is extracted from the simulation. From the values of the elements of this strain tensor ($\epsilon_{xx}, \epsilon_{yy}, \epsilon_{xy}$) we can compute the biaxial and uniaxial components of the strain at each point using: 

\begin{equation}
    \epsilon_{bi}=\frac{\epsilon_{xx}+\epsilon_{yy}}{2}
    \label{hyd}
\end{equation}

and 

\begin{equation}
    \epsilon_{uni} = \sqrt{ \Biggl( \frac{\epsilon_{xx}-\epsilon_{yy}}{2} \Biggl)^2+\epsilon_{xy}^2}
    \label{dev}
\end{equation}

\noindent as explained in the SI. Note that the uniaxial component of the strain ($\epsilon_{uni}$) is zero when $\epsilon_{xx} = \epsilon_{yy}$ and $\epsilon_{xy}=0$. Fig.~\ref{hyd_ps_strain}(a) and (b) plot the biaxial and uniaxial strains, respectively, as a function of position around a 12 $\mu$m ring covered by a 250 nm thick membrane of Ga$_2$Se$_2$. The position of the ring is indicated by the purple circle.

\subsection{Ring Structures: Biaxial Strain}
\begin{figure}[h]
\includegraphics[width=0.85\linewidth]{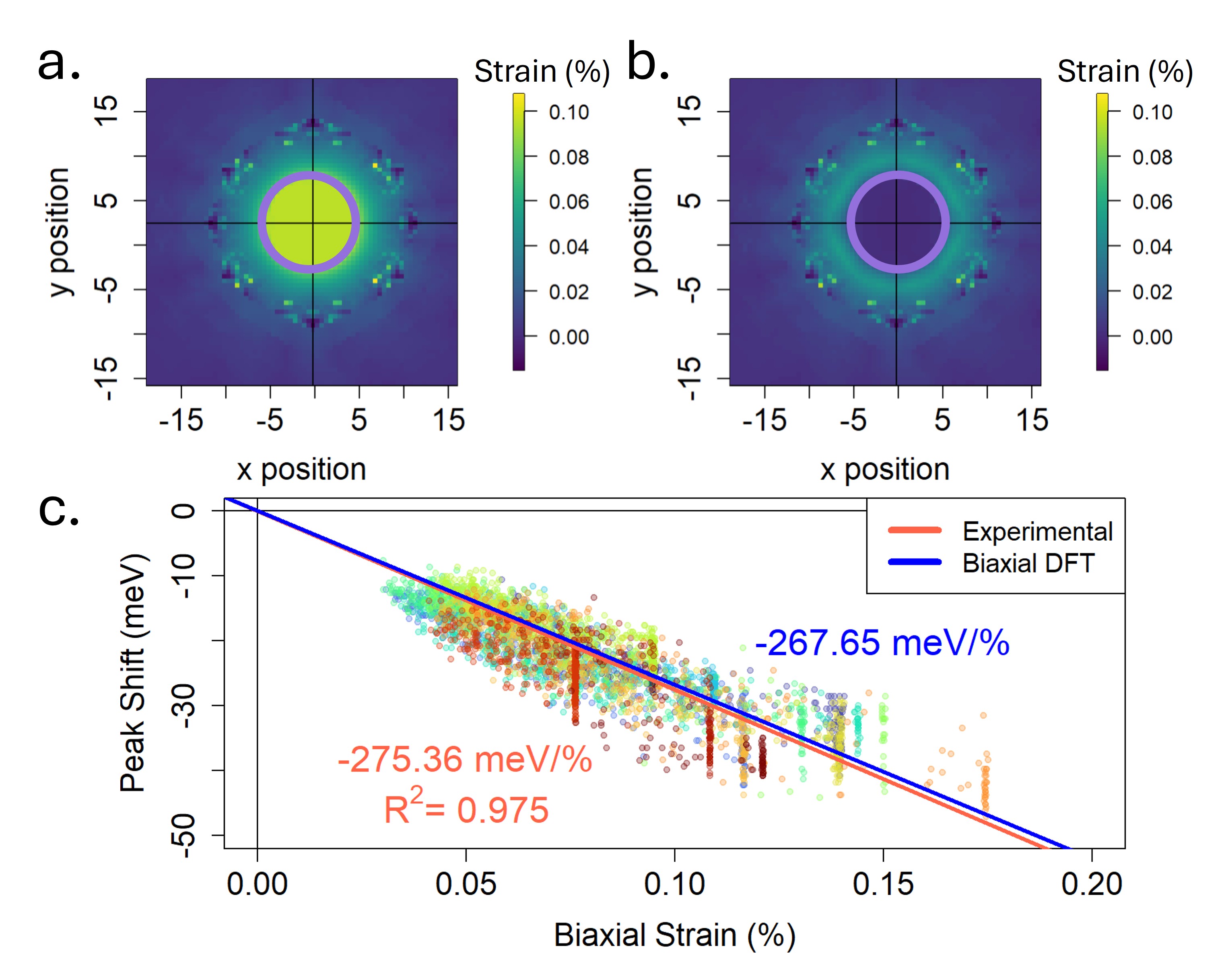}
  \caption{(a) biaxial strain and (b) uniaxial strain in one example ring structure. The purple ring indicates the position of the ring structure on the substrate. (c) The experimental peak shift as a function of biaxial strain from the suspended center region of 16 ring samples (4523 data points in total). The colors of the points indicate different samples. The fit to the experimental data is shown by the red line and the DFT results are shown by the blue line.}
  \label{hyd_ps_strain}
\end{figure}

We can now correlate the experimental measurements of PL peak shift with the computed local strain at each point. We begin by focusing on the area inside of each ring. As shown in Fig.~\ref{hyd_ps_strain}(b), the uniaxial strain within this central region of the ``drumhead" is negligibly small, which means restricting ourselves to data obtained from this region allows us to isolate the effects of biaxial strain. In Figure \ref{hyd_ps_strain}(c) we plot the experimentally measured peak shift as a function of computed biaxial strain. Each data point in Figure \ref{hyd_ps_strain}(c) corresponds to a unique PL spectra (4523 in total) obtained by studying 16 distinct ring structures (colors) with diameters ranging from 4 $\mu$m to 14 $\mu$m. The data are fit very well (R$^2$=0.975) by the red line with a slope (strain gauge factor) of $\beta_{exp}$=-275.4 meV / \% strain. We next compute the dependence of the band gap shift on biaxial strain using DFT (see Experimental Methods and Supporting Information). The biaxial strain gauge factor calculated using DFT is $\beta_{DFT}=-268$meV/\% strain, which is shown by the blue line in Figure \ref{hyd_ps_strain}(c) and agrees very well with the experimentally measured value (2.8\% error).

\subsection{Ridge Structures: Uniaxial Strain}
\begin{figure}[h]
  \includegraphics[width=.85\linewidth]{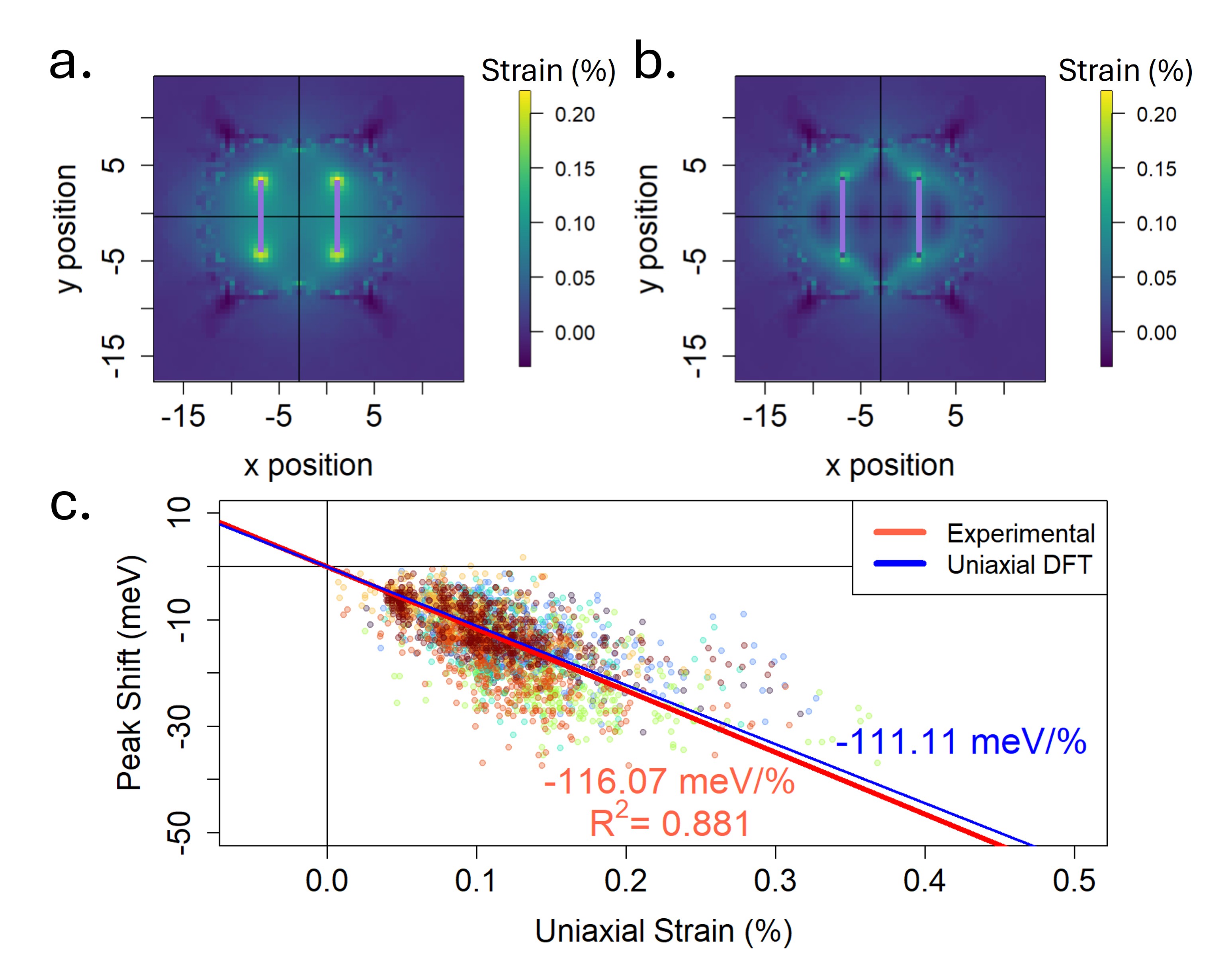}
  \caption{ (a) the biaxial and (b) uniaxial strain of a sample suspended over a ridge structure. The purple lines indicates the position of the ridge structure on the substrate. (c) The experimental peak shift as a function of uniaxial strain from 7 ridge samples (2803 data points in total). The colors of the points indicate different samples. The DFT results are shown in blue while the experimental fit is in red.}
  \label{dev_ps_strain}
\end{figure}

Figures \ref{dev_ps_strain}(a) and (b) plot the biaxial and uniaxial strain, respectively, for an example ridge structure with a separation distance of 12 $\mu$m and a Ga$_2$Se$_2$ flake thickness of 107 nm. We perform similar calculations for ridges separated by a distance ranging from 6 $\mu$m to 14 $\mu$m and covered by flakes ranging in thickness from 75 nm to 108 nm. We then numerically screen the computed strain at every point on every sample to identify the subset of locations for which the principle components of strain ($\epsilon_1$ and $\epsilon_2$, see Supporting Information) have a ratio less than or equal to 0.15. Because this ratio is below the Poisson Ratio of 0.23 \cite{mech_props}, such locations are areas of nearly pure uniaxial strain. Following the procedure described above, in Figure \ref{dev_ps_strain}(c) we use this subset of the data to plot the experimentally measured PL peak shift as a function of uniaxial strain with 2803 data points obtained from measurements of 7 distinct ridge structures. The data are fit very well (R$^2$=0.8812) by the red line with a slope of $\alpha_{exp}$= -116.1 meV / \% strain. We again compare the experimentally measured peak shift to DFT calculations. The uniaxial strain gauge factor calculated using DFT (see Experimental Methods and Supporting Information) is $\alpha_{DFT, z}$ = -111 meV / \% strain when strain is applied along the zigzag crystal axis and $\alpha_{DFT, a}$ = -108 meV/ \% strain when strain is applied along the armchair crystal axis. $\alpha_{DFT, z}$ = -111 meV / \% strain is shown by the blue line in Figure \ref{dev_ps_strain}(c) and has 4.4\% error relative to $\alpha_{exp}$. As described in the SI, we do not observe any change in the experimentally measured PL peak shift as a function of strain when strain is applied at different angles relative to the crystal axes, which is consistent with the nearly identical values for $\alpha_{DFT, a}$ and $\alpha_{DFT, z}$ obtained from DFT. 

\subsection{Complex Strain Profiles}
\begin{figure}[h]
  \includegraphics[width=1\linewidth]{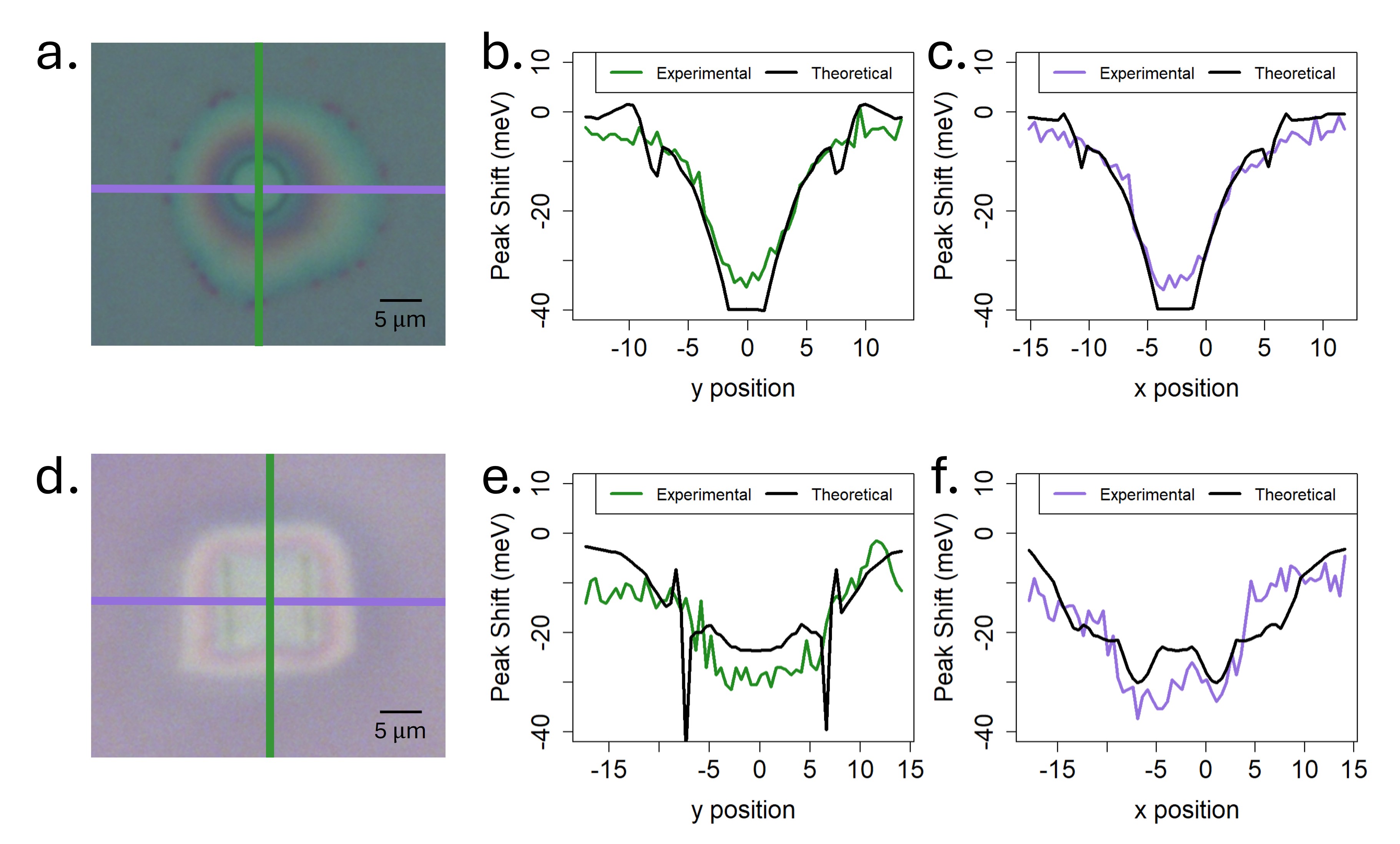}
  \caption{(a,d) Optical images of ring and ridge structures. The experimentally-measured PL peak shift along the green cross sections are plotted in (b,e) and those along the purple cross sections are plotted in (c,f). In all cases the black lines show the PL peak shift computed using Equation \ref{PL peak shift}.}
  \label{cross_section}
\end{figure}

By focusing on subsets of our data we extracted experimental measures of both the biaxial (isotropic) and uniaxial (deviatoric) strain gauge factors for thin Ga$_2$Se$_2$. We can now combine these results to understand the PL peak shifts in spatial locations that can not be described as nearly purely uniaxial or biaxial. The change in energy at each spatial location can be approximated by

\begin{equation}
    \Delta E \approx \beta_{exp}\frac{\epsilon_1+\epsilon_2}{2}+\alpha_{exp}\frac{|\epsilon_1-\epsilon_2|}{2}
    \label{PL peak shift}
\end{equation}

\noindent where $\beta_{exp}$ and $\alpha_{exp}$ are the experimentally determined strain gauge factors and $\epsilon_{1}$ and $\epsilon_{2}$ are the local principal components of strain extracted from the COMSOL simulations \cite{strain_calc,strain_calc02}. For example, in Figure \ref{cross_section} we show the peak shift along the horizontal (Figure \ref{cross_section}(b,e)) and vertical (Figure \ref{cross_section}(c,f)) cross sections depicted in Figure \ref{cross_section}(a,d). The green and purple lines show the experimentally measured PL peak shift. The black lines show the PL peak shift computed with Equation \ref{PL peak shift} using the computed principle components of strain at each point. There is excellent agreement between the experimental and computed values. Figure \ref{cross_section}(f) provides a clear example of how the simple model described by Equation \ref{PL peak shift} can predict complex bandgap / PL shifts arising from complex strain profiles such as those between -5 and +5 $\mu$m.

\begin{figure}[h]
  \includegraphics[width=1\linewidth]{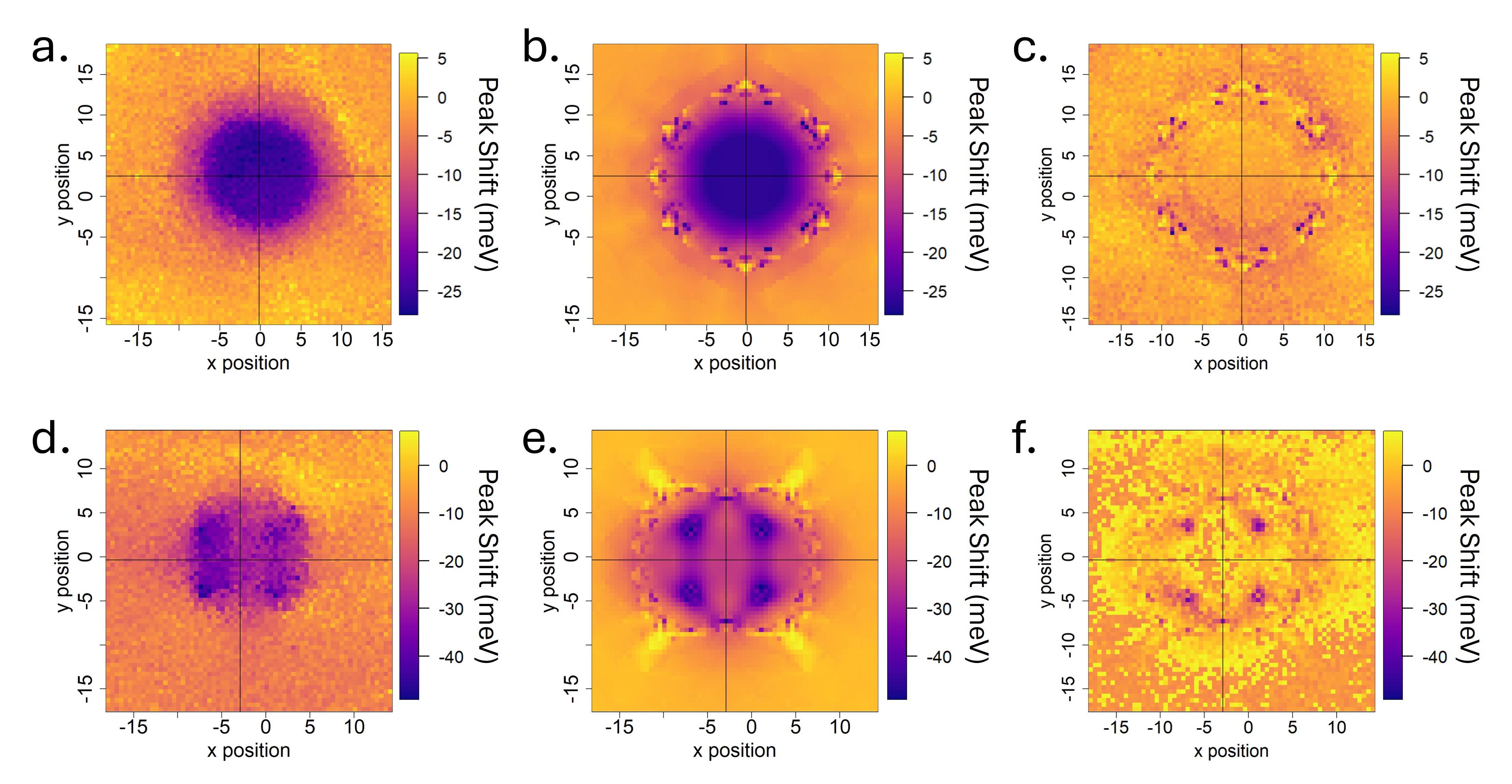}
  \caption{Maps of (a,d) the experimental peak shift, (b,e) calculated peak shift using experimentally determined strain gauge factors, and (c,f) the difference between the two for (a-c) a ring structure and (d-f)a ridge structure.}
  \label{delta}
\end{figure}

The cross sectional views presented in Figure \ref{cross_section} provide a clear picture of the excellent agreement between the experimentally-measured PL peak shift and the predictions of Equation \ref{PL peak shift}, but they are necessarily one-dimensional slices through the available data. In Figure \ref{delta} we provide two-dimensional plots of (a,d) the experimentally-measured peak shift, (b,e) the theoretical peak shift computed using Equation \ref{PL peak shift}, and (c,f) the difference between these two plots for (a-c) a ring of 12 $\mu$m diameter covered by a Ga$_2$Se$_2$ membrane of 259 nm thickness and (d-f) a ridge with separation distance of 8 $\mu$m and flake thickness of 75 nm (Figure \ref{delta}(d-f)). The near-zero values across most of Figures \ref{delta}(c) and (f) demonstrate that there is good agreement between the experimental and calculated maps of peak shift, with the main differences occurring at points of high strain (e.g.~tips of the ridges) or where we see numerical artifacts in the COMSOL simulations (e.g.~where the flake touches the substrate). 

Looking across all 23 structures measured here, we find a normalized root mean square error (RMSE) of 11.29\% between the experimental measurements and the predictions of Equation \ref{PL peak shift}. The RMSE is 10.15\% if we consider only ring structures and 13.88\% if we consider only ridge structures (see Supporting Information). To understand the primary source of this error we compute the normalized RMSE values between the height profile of each structure as determined experimentally by AFM with that extracted from the COMSOL simulations. We find that for all 23 structures the normalized RMSE in height is 7.22\% (7.25\% for only rings and 7.16\% for only ridges). In other words, the majority of the error between the experimentally-measured PL peak shift and that predicted by Equation \ref{PL peak shift} is due to the fact that the COMSOL simulations cannot predict the small local variations in flake height and strain that are unavoidable in any experiment. This suggests that Equation \ref{PL peak shift} can predict the PL peak shift with an error less than 10\% provided the local strain is known precisely.

\section{Conclusions}
%summarize\\
In summary, we developed a method to reproducibly induce strain on thin Ga$_2$Se$_2$ flakes exfoliated onto nanofabricated substrates and to compute the resulting local strain using simulations fully constrained by AFM measurements of the flake deformation profile. By correlating more than $10^3$ experimentally-measured PL peak shifts with the computed biaxial and uniaxial components of strain, we extract experimental values for the biaxial ($\beta_{exp}$=-275.4 meV / \% strain) and unixial ($\alpha_{exp}$ = -116.1 meV / \% strain) strain gauge factors. These experimental values are in excellent agreement with those we compute by DFT. Finally, we show that the PL spectra obtained when both biaxial and uniaxial strain are present can be fully explained using a simple model based only on the experimentally-obtained values for the strain gauge factors. The results provide a statistically-meaningful, comprehensive, and experimentally-validated understanding of the relationship between applied strain and local shifts in the bandgap of the 2D material Ga$_2$Se$_2$, which in turn provides a theoretical framework for designing strain profiles to achieve spatial control over bandgap. The methods reported here can be immediately applied to any direct bandgap 2D material and provide a framework for the design of systems with predictable strain profiles and bandgap shifts. This predictable control over strain in 2D materials will advance photonics technologies by, for example, advancing the realization of deterministic arrays of quantum emitters.

\section{Methods}
\subsection{Sample Preparation}
%nanofab and exfoliation (around 2 paragraph) \\
%mention structures 
The patterned substrates were fabricated in the University of Delaware's Nanofabrication Facility (UDNF). The nanostructures were fabricated on silicon-on-insulator (SOI) substrates using methods adapted from a previously reported process outlined in more detail in the Supplemental Information \cite{Herrmann2024}. Once samples were cleaned and a bilayer resist stack was applied, the resist was patterned using electron beam lithography (Raith EBPG 5200, 100kV). A chromium (Cr) hard mask was then deposited by electron beam evaporation (PVD Products) and later removed with etchant. The nanostructures were etched into the substrate using an inductively coupled plasma etching system. 

The $\epsilon$-Ga$_2$Se$_2$ bulk crystal used was purchased from 2D Semiconductors. Large area flakes were exfoliated from this bulk crystal using polydimethylsiloxane (PDMS) strips (Gel-Pak). The flakes were then deterministically transferred onto the fabricated substrates through a water-assisted procedure (see Supplemental Information) \cite{Li2015}. To compensate for the unavoidable optical degradation that occurs after exfoliation and transfer, which we have previously reported\cite{degradation}, samples were held for 24 hours before PL maps were obtained.

\subsection{Measuring PL}
%Methods/temp/xy raster\\
%each structure has \# of pixels\\
%PL peak extraction\\

A Horbia LabRAM HR Evolution bench top microscope was used for all PL collection. Measurements were taken at room temperature (298K) and were performed using an excitation wavelength of 532 nm with an intensity of 1.7 mW at the surface of the sample. Each optical measurement was performed using a 50X objective (1 $\mu$m spot size), dispersed with a 300 groove/mm grating to accumulate spectra spanning 575-725nm (1.71-2.16 eV), and measured with a Silicon CCD array detector with a 0.1 s integration time. For each structure, a map was taken over the entire strained area. The number of points in each map varied between 55x55 points to 80x80 points, depending on the size of the feature. The step size of the map was consistently 0.5 $\mu$m in both the x and y direction. 

All data analysis was done using RStudio running R version 4.3.2 using the packages `car',  `stats', `fields', `mclust', `viridis', and `RColorBrewer' \cite{car,stats,fields, mclust, viridis, brewer}. Raw data was imported from text files extracted from Horbia's LabSpec 6 Spectroscopy Suite during collection. All PL data was normalized to unit area and the baseline was subtracted.

\subsection{Atomic Force Microscopy}
Atomic Force Microscopy (AFM) was conducted with an Anasys Nano IR2 in the Advanced Materials Characterization Laboratory (AMCL) at the University of Delaware. The AFM scans were first conducted on the edge of the flake to determine the flake thickness, which is used as an input for the COMSOL simulations. The thickness of flakes was determined by averaging the height profile across a 100 pixel area using the Gwyddion software package. Full AFM scans of every suspended flake provide the full deformation profile that was used to constrain the boundary load for each COMSOL simulation. The height of the bare nanostructures (i.e.~rings, ridges) was also confirmed via AFM.

\subsection{Quantifying Strain}
%overview (1 paragraph) \\
%comsol simulation method (1-2 paragraph)\\
%- unique flake thickness in each calc \\
%- 1 fitting parameter validated against AFM profile \\
%results\\

The finite element method in COMSOL Multiphysics was used to perform strain simulations and estimate the local strain components (see Supplemental Information). The Ga$_2$Se$_2$ flakes were modeled as a membrane with biaxial mechanical properties, a Young's modulous of 82 GPa, and Poisson's ratio of 0.22 \cite{Chitara2018}. The flake thickness and nanostructure height obtained via AFM were also used as inputs to the simulation. To simplify the simulations, the deformation behavior of the Ga$_2$Se$_2$ flakes were approximated by applying a boundary load in the direction of the substrate. Deformation profiles were extracted for various boundary loads and compared to profiles obtained via AFM using a least-squared approach (see Supplemental Information). Once the appropriate boundary load is determined, the simulation results are exported from COMSOL. These results are over a spatial grid 30 $\mu$m by 30 $\mu$m with 0.1 $\mu$m spacing. To directly compare to experimental results we apply a Gaussian smoothing filter (using the isoblur function in R with a sigma value of 5) and trimmed and centered the results to match the spatial coordinates of experimental maps. This process was performed using RStudio (version 4.3.2) using the packages `soundgen', `pracma', `imager', and `akima' \cite{soundgen,pracma,imager, akima}. 

\subsection{DFT calculations}
Structural parameters for DFT calculations were determined using the PBEsol exchange-correlation functional,\cite{SI_dft01} while the electronic properties were calculated with the Heyd-Scuseria-Ernzerhof (HSE) screen hybrid functional including spin-orbit coupling (SOC). All simulations were conducted using the VASP code with projector augmented plane wave (PAW) potentials, a plane-wave cutoff of 400 eV, and a 9×9×1 Monkhorst-Pack k-point mesh. Strain was applied along both the armchair (y) and zigzag (x) directions in the plane of the crystal. For each case, the lattice was adjusted to maintain zero stress in the transverse direction, and the out-of-plane lattice parameter was fully relaxed. The band structure was computed for strains ranging from 0\% to 5\%. Results show a linear decrease in band gap with increasing uniaxial strain: -0.108 eV/\% for strain along the armchair direction and -0.111 eV/\% along the zigzag direction. Importantly, the material retains its direct band gap across the full strain range, with the conduction and valence band edges remaining at the $\Gamma$ point. 
%keep sentance below for future use
%These simulations provide a theoretical basis for interpreting the strain induced redshift and enhanced photoluminescence observed experimentally. The increased separation of nearby band edges under strain suggests an even more direct gap character, which may contribute to the observed intensity increase in PL measurements.

\section{Conflict of Interest}
EH and XW are listed as inventors on a pending patent application by the University of Delaware.

\begin{acknowledgement}
LLM and MFD gratefully acknowledge support from the National Science Foundation (Grant 2217786) and the II-VI / Coherent Foundation through their Block-Gift Program. EH and XW acknowledge the support from the U.S. National Science Foundation under Grant DMR-2128534. The authors also acknowledge the use of equipment in the Surface Analysis Faculty at UD supported by the National Science Foundation (Major Research Instrumentation Award Number: CHE-1828325).
\end{acknowledgement}

\begin{suppinfo}

\noindent The Supporting Information includes detailed explanation of DFT calculations, COMSOL Multiphysics simulations, sample preparation, PL hyperspectral imaging, and data analysis.  
\end{suppinfo}


\bibliography{ref}

\end{document}

\section{Density Functional Theory (DFT) Calculations}
%Igor \\
%1-3 pages about DFT calculations \\
%\textcolor{red}{IGOR/ ANDERSON - PLEASE REVIEW THIS SECTION CAREFULLY - I MADE EDITS THAT NEED TO BE VERIFIED}

Our density functional theory (DFT) calculations are based on the PBEsol exchange-correlation functional for determining the structural parameters \cite{SI_dft01}. The interactions between valence electrons and ionic cores were represented using the projector augmented plane wave (PAW) method as implemented in the VASP code \cite{SI_dft03}. The $\epsilon-$Ga$_2$Se$_2$ model was constructed with the valence electronic configurations of Ga and Se atoms being 4s$^2$4p$^1$ and 4s$^2$4p$^4$, respectively. The electron wave functions were expanded in plane waves with an energy cut-off of 400 eV and the Brillouin zone was sampled with an 9x9x1 Monkhorst-Pack grid of the k points. We fully optimized all geometric structures until the convergence of total energy and force reached 10$^{-7}$ eV and 10$^{-4}$ eV\AA$^{-1}$, respectively. 

\begin{figure}
  \includegraphics[width=0.75\linewidth]{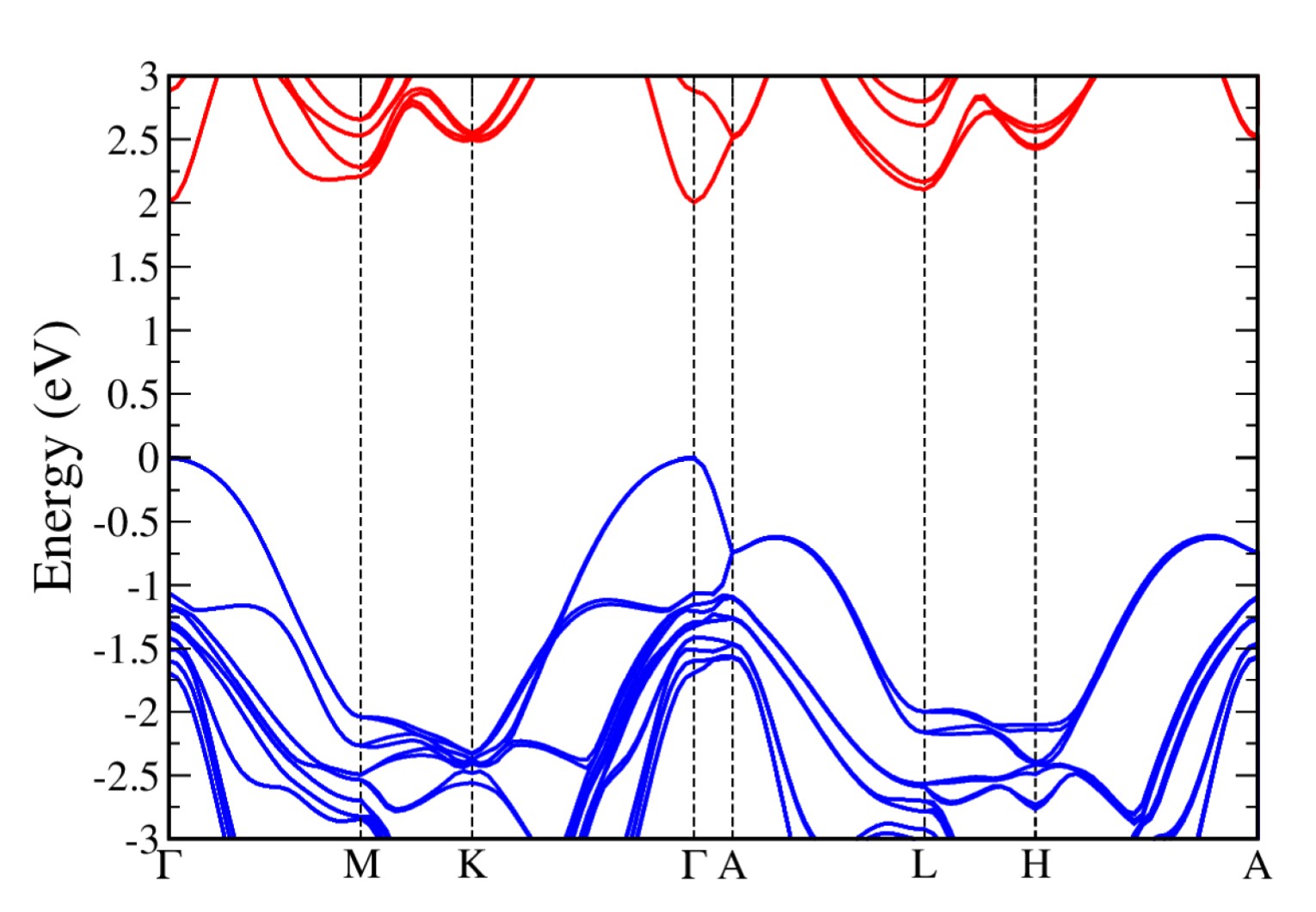}
  \caption{Band structure of $\epsilon-$Ga$_2$Se$_2$ using HSE+SOC functional. The band gap is 2 eV and located at the $\Gamma$ point}
  \label{SI_dft_bandstructure}
\end{figure}

We next used the Heyd-Scuseria-Ernzerhof (HSE) screened hybrid functional with a mixing parameter $\alpha=27\%$ to calculate the band structures \cite{SI_dft02}. We again used a 9x9x1 Monkhorst-Pack k-point grid and included spin-orbit coupling (SOC). Figure S\ref{SI_dft_bandstructure} shows the electronic band structure of $\epsilon-$Ga$_2$Se$_2$ calculated using the HSE+SOC functional. The calculations predict a direct band gap of 2 eV at the $\Gamma$ point, which agrees well with available experimental data.\cite{SI_dft05} This agreement highlights the reliability of hybrid functionals for accurately capturing the electronic properties of Ga$_2$Se$_2$ and tells us that HSE+SOC functionals should be used for extracting the bandgap as a function of applied strain.

\begin{figure}
  \includegraphics[width=0.75\linewidth]{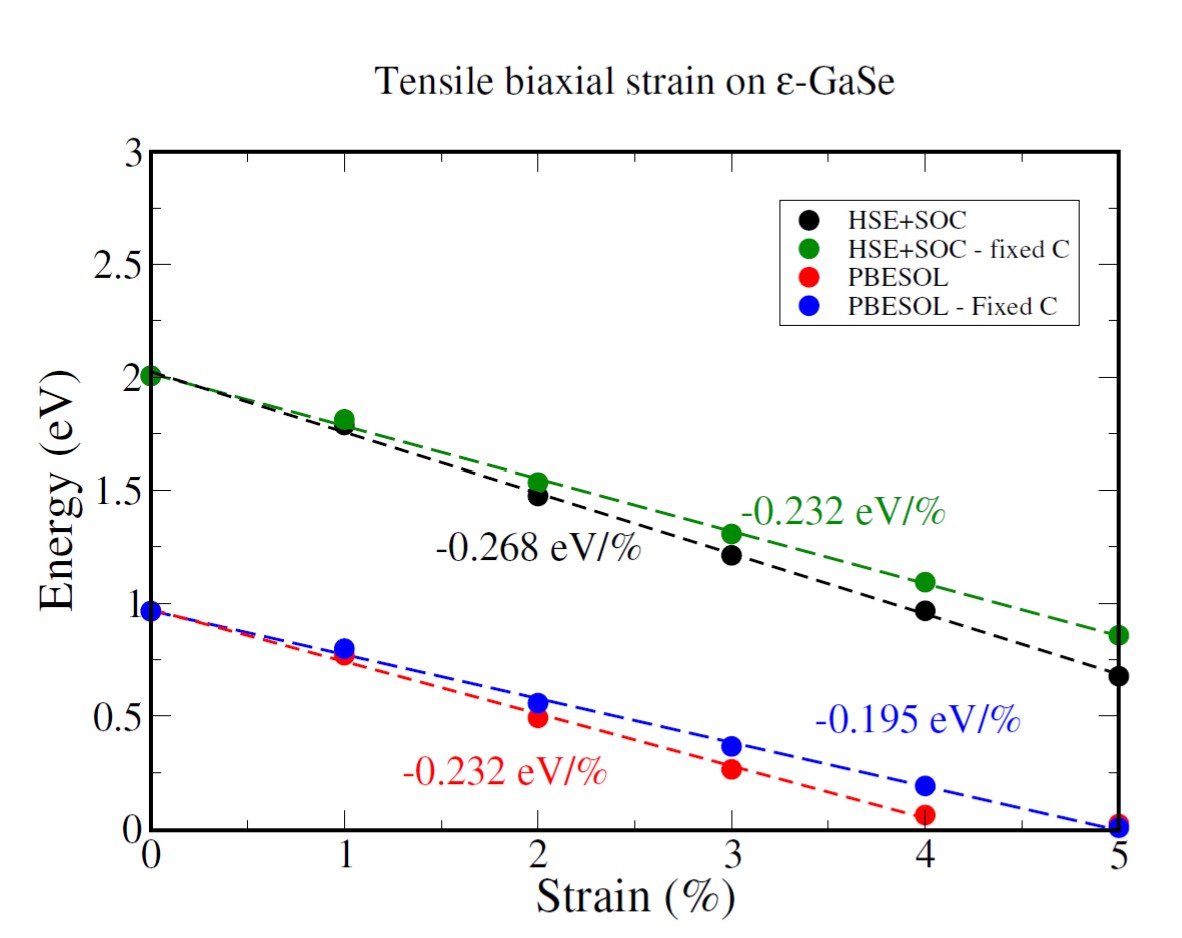}
  \caption{Band gap evolution with applied biaxial strain in Ga$_2$Se$_2$ using different functionals}
  \label{SI_dft_biaxial}
\end{figure}

Fig.~S\ref{SI_dft_biaxial} shows the change in bandgap energy computed via DFT as a function of applied biaxial strains ranging from $0\%$ to $5\%$. The blue and red data points reflect the results of calculations using the PBESOL functional with (red) and without (blue) allowing the relaxation of atoms along the c axis (perpendicular to the applied biaxial strain). Because PBE functionals typically underestimate band gaps, we then repeat these DFT calculations using the HSE+SOC functional; the results are shown by the black and green data points again with (black) and without (green) allowing relaxation along the c axis. Because 1) the HSE+SOC functional is known to better predict bandgap energies and 2) relaxation of the lattice along the c axis is expected, we take -0.268 eV / \% strain as the value for the biaxial strain gauge factor reported in the main manuscript.

We applied uniaxial stress along the armchair (y) direction by setting the y component of the lattice vectors to a specific axial strain and adjusting the x component to find the transverse strain that minimizes the total energy. Similarly, for stress along the zigzag (x) direction, we set the x component to a given axial strain and varied the y component to determine the optimal transverse strain. Uniaxial strain is applied along a single in-plane direction, while the transverse in-plane and out-of-plane directions are allowed to relax. All these directions and the atomic positions were relaxed using the PBEsol level of theory. 

\begin{figure}
  \includegraphics[width=0.75\linewidth]{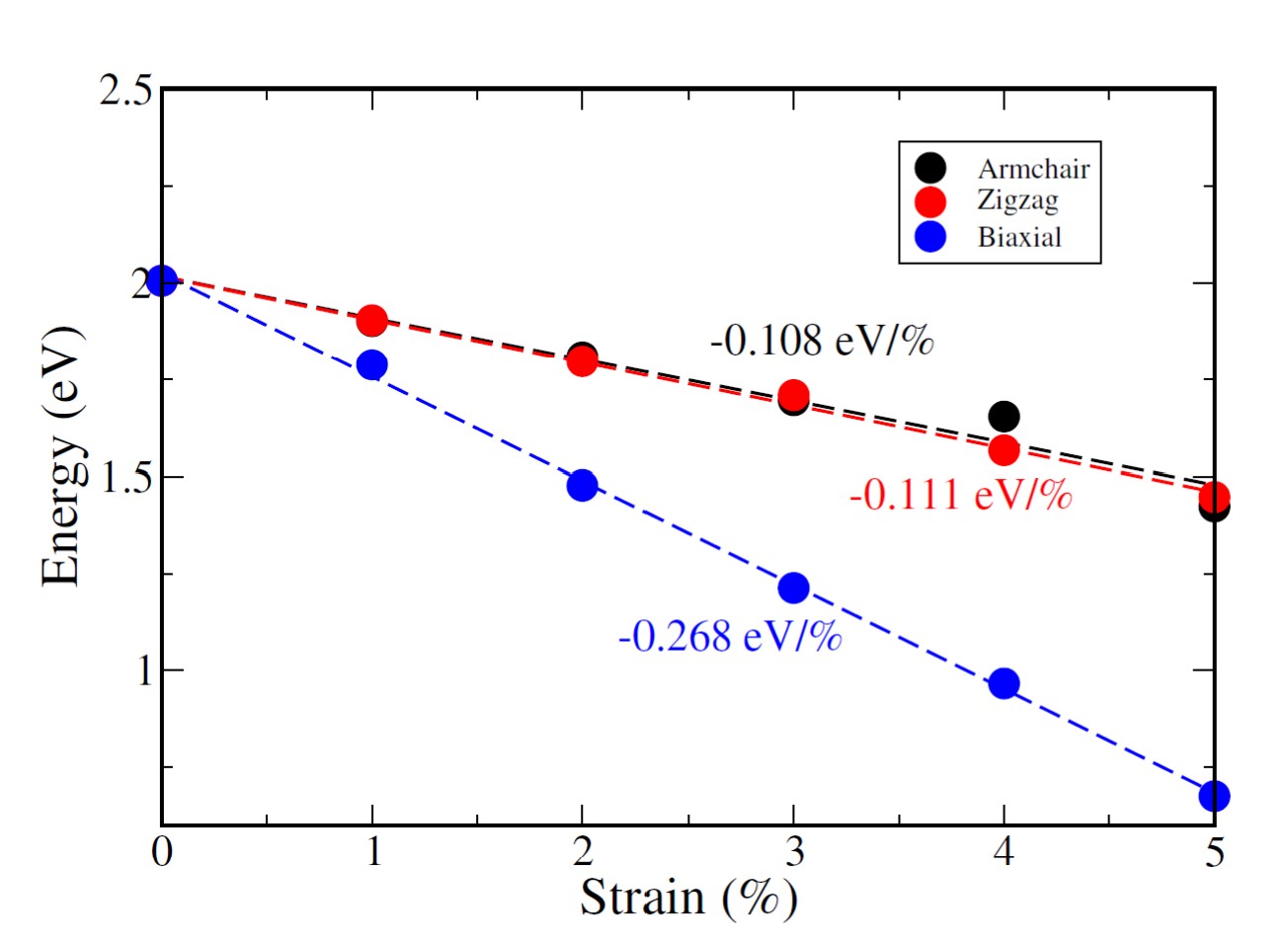}
  \caption{Band gap evolution of Ga$_2$Se$_2$ with tensile strain in armchair (black) zigzag (red) direction and biaxial strain. The HSE+SOC functional was used.}
  \label{SI_dft_tensile}
\end{figure}

\begin{figure}
  \includegraphics[width=0.75\linewidth]{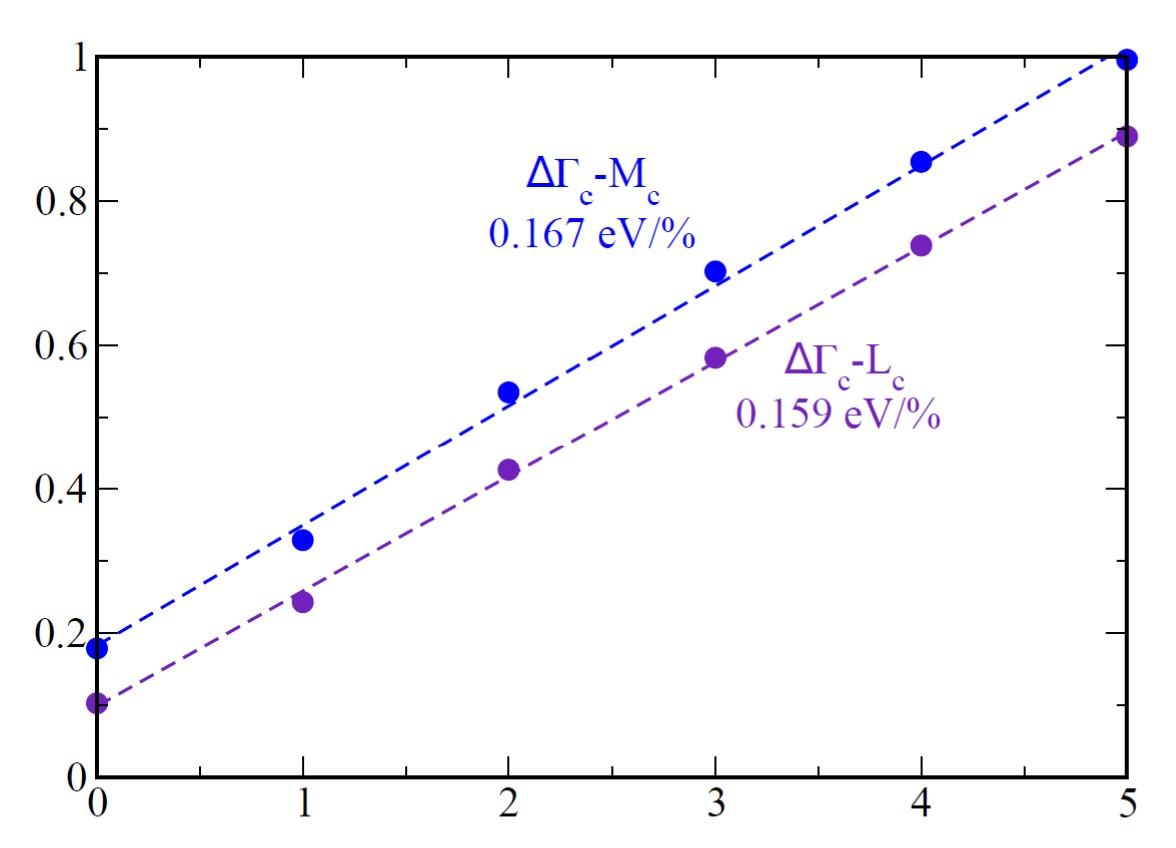}
  \caption{Energy difference between $\Gamma$ and L symmetry points (blue) and $\Gamma$ and M (purple) in the conduction band}
  \label{SI_dft_conduction}
\end{figure}

As illustrated in Figure S\ref{SI_dft_tensile}, the band gap energies decrease linearly with applied uniaxial tensile strain. The rate of decrease is -0.108 eV$ / \%$ under uniaxial strain in the armchair direction and $-0.111$ eV$ / \%$ in the zigzag direction. Throughout the range of modeled strain, Ga$_2$Se$_2$ maintains its direct bandgap; no strain induced transitions from direct to indirect bandgaps nor band edge shifts are observed in the simulation results, in contrast to transition metal dichalcogenides (TMDCs) \cite{SI_dft04}. As shown in Figure S\ref{SI_dft_conduction}, the band edges at the M and L points increase in energy relative to the $\Gamma$ point. This means that the material has an even more direct bandgap when strain increases, which may explain why the PL intensity increases in strained regions. 

\section{Sample Preparation}
All nanostructures were fabricated on commercially available silicon-on-insulator (SOI) substrates using methods adapted from a previously reported process \cite{Herrmann2024}. The substrates were first cleaned with an acetone and isopropyl alcohol (IPA) wash. A bilayer resist stack consisting of lift-off (LOR3A) and electron beam (AR-P 6200.04) resists with a combined thickness of ~250 nm was then spun onto the bare SOI surface. The resist was patterned using electron beam lithography (Raith EBPG 5200, 100kV) and subsequently developed in AR 600-546 and IPA for the lift-off and electron beam resists, respectively. Rings and ridges of various diameters and separations were defined, each with an in-plane thickness of approximately 100 nm. A 25 nm thick chromium (Cr) hard mask was then deposited at 0.2 Å/s by electron beam evaporation (PVD Products). Following liftoff, the nanostructures were etched into the SOI substrate using an inductively coupled plasma etching system with CHF$_3$ (30 sccm) and C$_4$F$_8$ (40 sccm)  gases. The etch time was set to 10 minutes, resulting in a nanostructure height of approximately 420 nm. The remaining Cr hard mask was removed by immersing the samples in commercially available Cr etchant held at 40°C for one hour. 

Following the fabrication of the nanostructures, large-area $\epsilon-$Ga$_2$Se$_2$ flakes of relatively uniform thickness were exfoliated from a bulk crystal (2D Semiconductors) using polydimethylsiloxane (PDMS) strips (Gel-Pak). Exfoliated flakes were then deterministically transferred onto the patterned substrates through a water-assisted procedure \cite{Li2015}. A bead of water is first picked up at the location of the flake and then aligned above the target nanostructure. The flake is then lowered and brought into contact with the substrate at room temperature. With the flake and substrate in contact, the substrate is gradually heated to 95°C to enable the release of the flake from the PDMS and evaporation of the water. The sample is then allowed to cool to room temperature before performing photoluminescence mapping.

\section{Finite Element Simulations}
We employ the finite element method in COMSOL Multiphysics to perform all strain simulations presented in the main text. In all simulations, Ga$_2$Se$_2$ is modeled as a membrane with isotropic mechanical properties, Young’s modulus of 82 GPa, and Poisson’s ratio of 0.22 \cite{Chitara2018}. The thickness of the Ga$_2$Se$_2$ flake and height of the nanostructures are defined using experimentally measured values. The presence of the nanostructure is approximated by projecting its in-plane geometry onto the Ga$_2$Se$_2$ membrane surface. The projected region is constrained to in-plane motion to mimic frictionless contact between the Ga$_2$Se$_2$ flake and the apex of the nanostructure. 

The experimental transfer procedure involves complicated interactions between the Ga$_2$Se$_2$ flake and nanostructured surface, including deformation of the PDMS stamp and capillary forces from the interaction with water. To simplify the simulations, we approximate the deformation behavior of the Ga$_2$Se$_2$ flake by applying a boundary load to the upper surface of the Ga$_2$Se$_2$ in the direction of the substrate. Contact between the Ga$_2$Se$_2$ and flat substrate surface is defined such that the Ga$_2$Se$_2$ domain adheres to the substrate once contact is made. To limit the number of fitting parameters, the simulations exclude frictional forces and decohesion of the Ga$_2$Se$_2$ from the substrate. The only free parameter in the fit is the boundary load, which is individually calibrated for each structure by varying the magnitude of the force per unit area and comparing the resulting simulated deformation profiles to the measured AFM height profiles using a least-squares approach. 

\begin{figure}
  \includegraphics[width=1\linewidth]{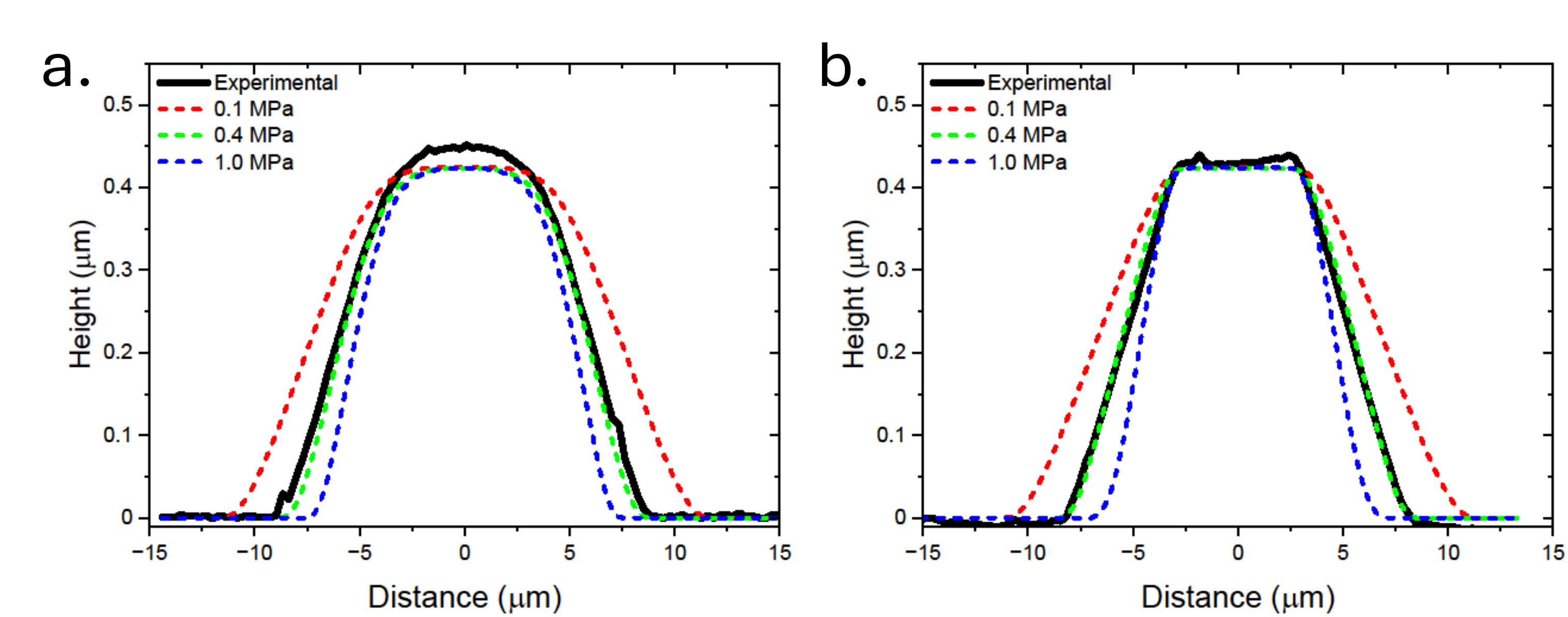}
  \caption{Comparisons of experimental and simulation height profiles taken along the (a) horizontal and (b) vertical cross-sections for a Ga$_2$Se$_2$ flake transferred onto a representative ridge nanostructure.}
  \label{si_ridge_calibration}
\end{figure}

The calibration procedure was performed on all nanostructures and an optimized boundary load was used for each individual structure. These range from 0.1 MPa and 0.4 MPa. In Figure S\ref{si_ridge_calibration}, we compare simulated and AFM cross-sectional height profiles of Ga$_2$Se$_2$ transferred onto representative ridge nanostructures under various boundary loads. 

\begin{figure}
  \includegraphics[width=0.55\linewidth]{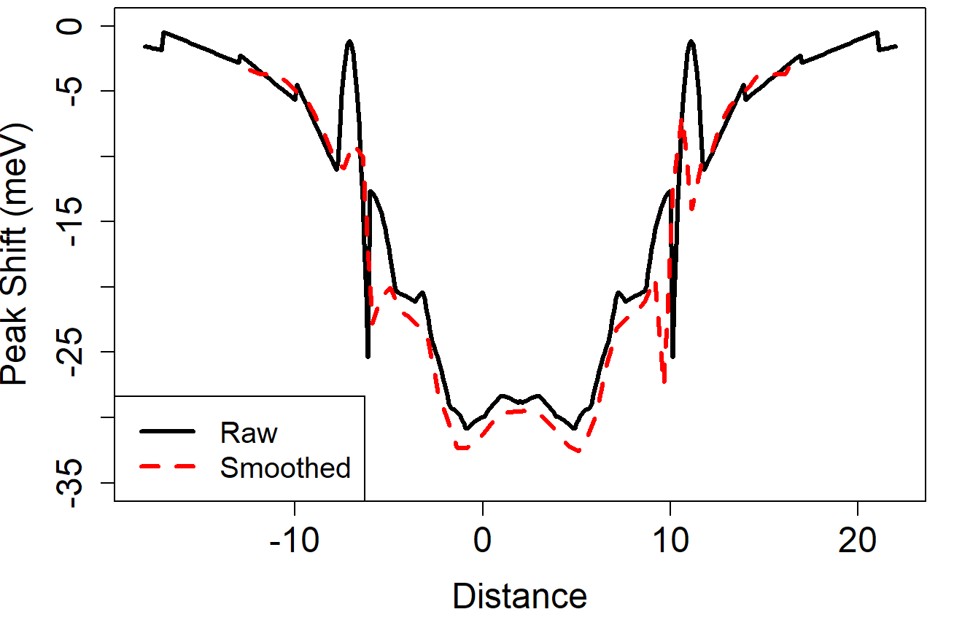}
  \caption{Comparison between raw (black) and smoothed (dashed red) simulated PL shift data taken along a cross-section through a Ga$_2$Se$_2$ flake suspended on a representative ridge nanostructure.}
  \label{si_shift}
\end{figure}

The strain components are extracted from the deformed Ga$_2$Se$_2$ simulation domain over a square grid of 30 µm width and point spacing of 0.1 µm. The finite nature of the simulation mesh elements introduces sharp transitions in strain magnitude and direction that are not captured by experimental photoluminescence maps. For direct comparison with experimental results, we apply a Gaussian smoothing filter of 2 µm diameter to the extracted simulation data to approximate the averaging effects of the laser spot size on the acquired data. We present an example of this smoothing process on the simulation data in Figure S\ref{si_shift}, where we show simulated PL shift data both before and after smoothing.

\section{Photoluminescence (PL) Measurements}
%include additional examples of other structures \\
%include figure similar to Fig 1 from main text but for ridges \\
%include figure with PL near rings and far from rings for peakshift subtraction\\

\begin{figure}
  \includegraphics[width=1\linewidth]{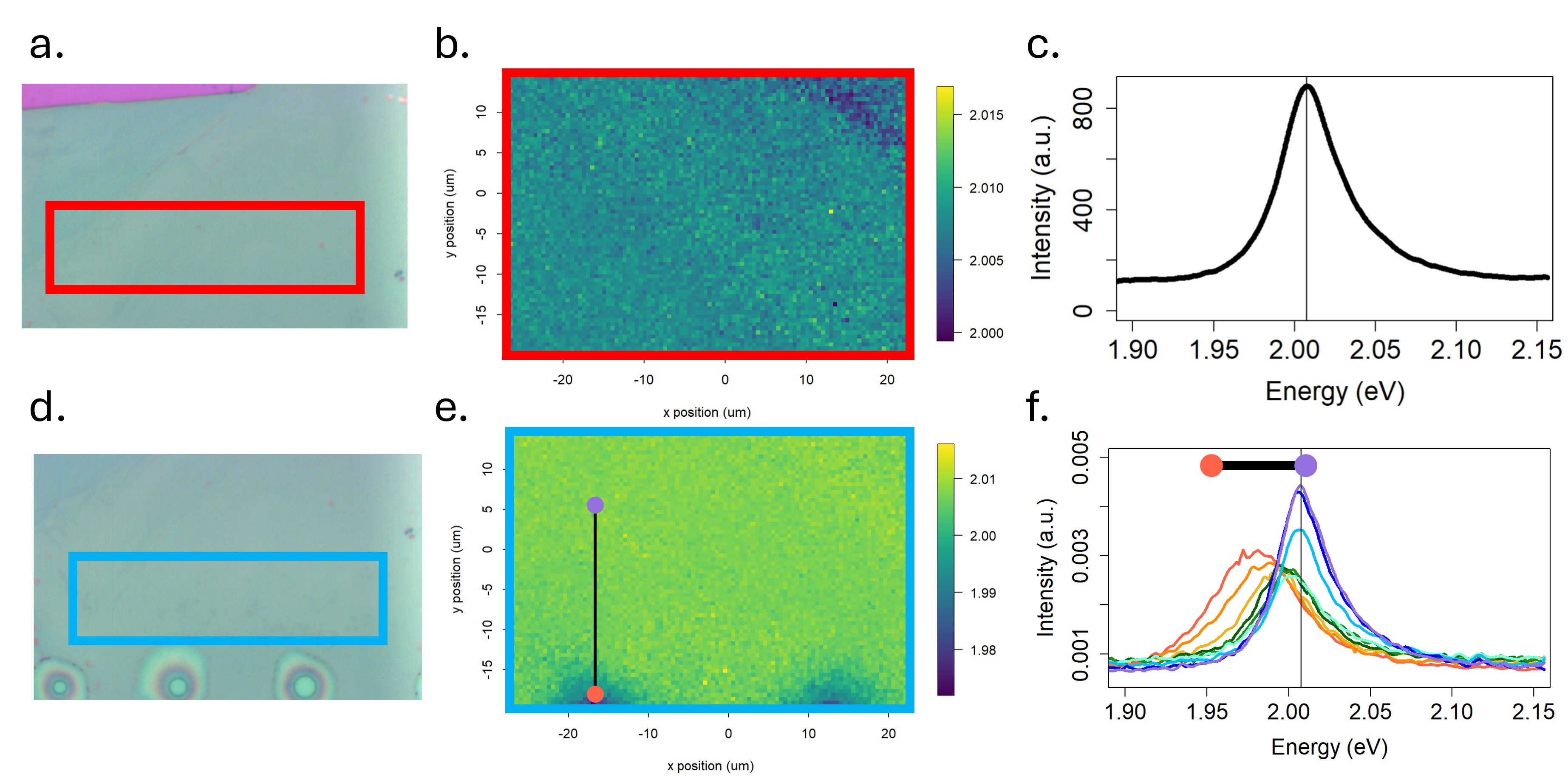}
  \caption{(a,d) Show the optical image of the measured area on a flat region of the flake that is suspended over the patterned substrate, with the resulting peak max energy shown in (b,e) for each spatial location. (c) Average PL signal across the entire area shown in (b). (f) Cross sectional view of the peak shift in (e) as the distance from the structure increases.}
  \label{flat}
\end{figure}

To determine the experimental PL peak shift, the baseline (unstrained) PL energy was first obtained from areas far away from any nanostructures, but within the same flake. The rectangles in Figures S\ref{flat} (a) and (d) provide two examples of optical images of `flat' areas that were probed. Figures S\ref{flat}(b) and (e) show the PL center energy measured at each point within the corresponding rectangles. Figure S\ref{flat}(c) shows the average of all points in Figure S\ref{flat}(b), which provides a clear measure of the unstrained PL peak energy. Figures S\ref{flat}(e) and (f) illustrate the importance of obtaining the baseline (unstrained) PL peak energy from regions that are far away from any nanostructures: a clear shift in PL energy is observed as a function of position due to the rings that are near, but not within, the lower boundary of the rectangle shown in Figure S\ref{flat}(d). The unstrained PL energy was determined independently for all unique flakes. These values of the unstrained PL peak energy were then used to calculate the peak shift across the suspended regions. 

\begin{figure}
  \includegraphics[width=0.75\linewidth]{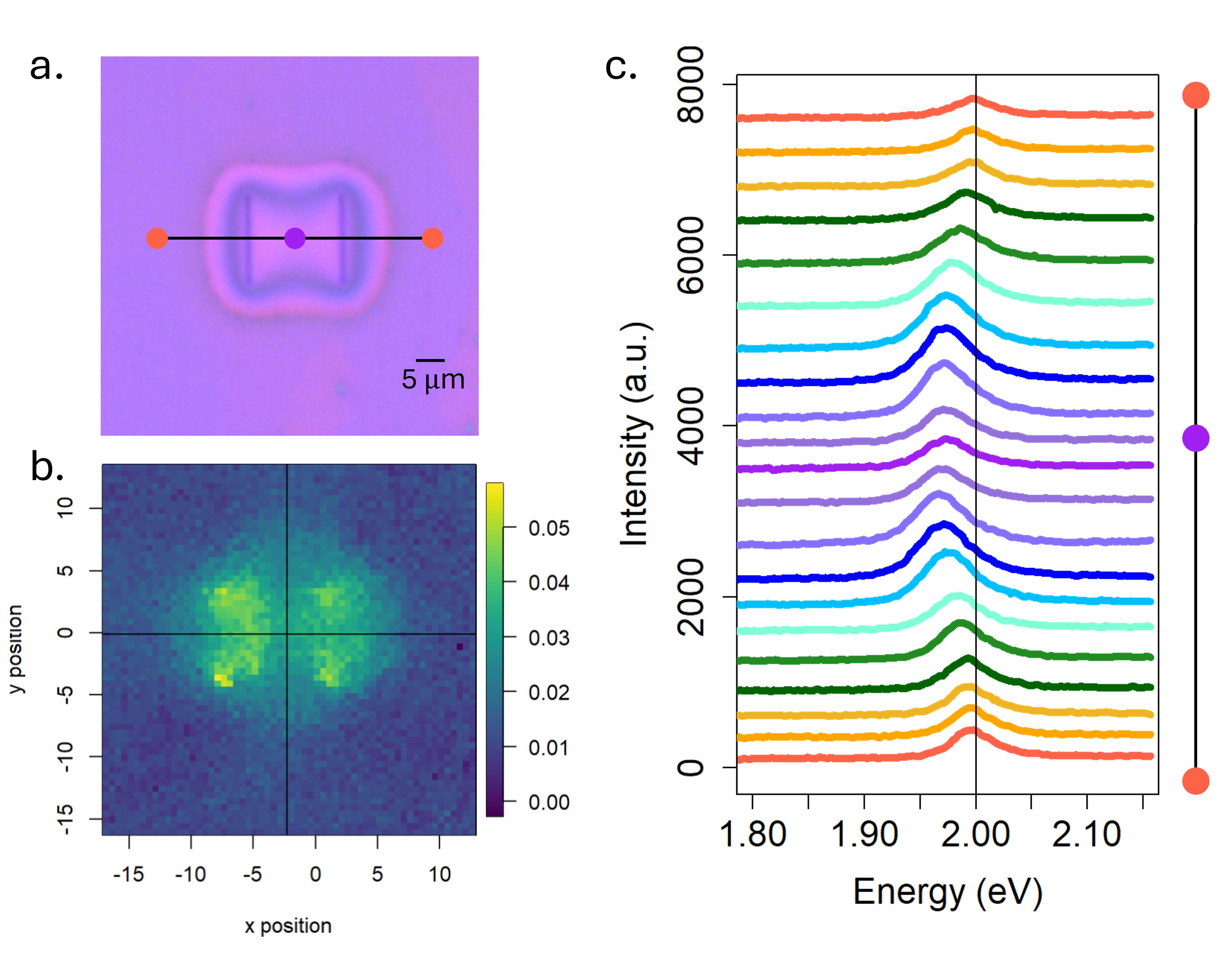}
  \caption{(a) Optical image of suspended flake over a ridge structure. (b) x-y peak shift map for this ridge structure. (c) Stacked spectra showing the peak shift along the line indicated in (a). }
  \label{SI_ridge_expPL}
\end{figure}

\section{Strain Components}
%\textcolor{red}{Eric / Xi check my revisions carefully, please.}
The 2D tensor that describes the local strain at any point is given by \cite{strain_calc,strain_calc02}:

\begin{equation}
    \epsilon = 
    \begin{pmatrix}
        \epsilon_{xx} & \epsilon_{xy} \\
        \epsilon_{xy} & \epsilon_{yy} \\
    \end{pmatrix}
    \label{strain_tensor}
\end{equation}

which can be broken into two components:

\begin{equation}
    \epsilon = 
    \begin{pmatrix}
        \epsilon_{xx} & \epsilon_{xy} \\
        \epsilon_{xy} & \epsilon_{yy} \\
    \end{pmatrix}
    =
    \begin{pmatrix}
        \frac{\epsilon_{xx}+\epsilon_{yy}}{2} & 0 \\
        0 & \frac{\epsilon_{xx}+\epsilon_{yy}}{2} \\
    \end{pmatrix}
    +
    \begin{pmatrix}
        \frac{\epsilon_{xx}-\epsilon_{yy}}{2} & \epsilon_{xy} \\
        \epsilon_{xy} & \frac{\epsilon_{yy}-\epsilon_{xx}}{2} \\
    \end{pmatrix}
    =
    \epsilon_{hyd}+\epsilon_{dev},
    \label{strain_tensor_decomposition}
\end{equation}

%\begin{equation}
%    \epsilon = 
%    \begin{pmatrix}
%        \epsilon_{xx} & \epsilon_{xy} \\
%        \epsilon_{xy} & \epsilon_{yy} \\
%    \end{pmatrix}
%    =
%    \begin{pmatrix}
%        \frac{\epsilon_{xx}+\epsilon_{yy}}{2} & 0 \\
%        0 & \frac{\epsilon_{xx}+\epsilon_{yy}}{2} \\
%    \end{pmatrix}
%    +
%    \begin{pmatrix}
%        \frac{\epsilon_{xx}-\epsilon_{yy}}{2} & 0 \\
%        0 & \frac{\epsilon_{xx}-\epsilon_{yy}}{2} \\
%    \end{pmatrix}
 %   =
 %   \epsilon_{bi}+\epsilon_{uni},
 %   \label{strain_tensor}
%\end{equation}

\noindent where $\epsilon_{hyd}$ is the hydrostatic component of the strain and $\epsilon_{dev}$ is the deviatoric component. It is straightforward to see that the hydrostatic component simply represents the application of biaxial strain, with equal strain along the $x$ and $y$ directions and no shear ($\epsilon_{xy}=0$). Mathematically, when $\epsilon_{xx}=\epsilon_{yy}$ and $\epsilon_{xy}=0$ the deviatoric component of the strain goes to zero and only the hydrostatic (biaxial) component remains. In other words the biaxial strain is given by

\begin{equation}
    \epsilon_{bi}=\frac{\epsilon_{xx}+\epsilon_{yy}}{2} = \frac{\epsilon_{1} + \epsilon_{2}}{2}
    \label{hyd}
\end{equation}

\noindent where $\epsilon_{1}$ and $\epsilon_{2}$ are the eigenvalues of the strain tensor (Equation \ref{strain_tensor})

\begin{equation}
    \epsilon_{1,2} = 
    \frac{\epsilon_{xx}+\epsilon_{yy}}{2}
    \pm
    \sqrt{ \Biggl( \frac{\epsilon_{xx}-\epsilon_{yy}}{2} \Biggl)^2+\epsilon_{xy}^2}.
    \label{principal_strain}
\end{equation}

\noindent known as the principal components. These principal components represent the maximum and minimum strain at a given spatial location. 

When we perform the biaxial strain gauge analysis reported in the main manuscript, we first numerically screen the computed strain to identify points at which the deviatoric component of strain is zero. We then use Equation \ref{hyd} to compute the local biaxial strain and plot the experimentally measured PL peak shift as a function of the biaxial strain. The fit to that data gives us the biaxial strain gauge factor $\beta_{exp}$.

We can see that both the maximum and minimum strains include the biaxial component. This makes sense because the biaxial strain is isotropic in the plane. If we subtract the isotropic biaxial component from each principal strain, the remainder describes any ``excess" strain along one dimension, i.e.~unixaial strain. In other words, the uniaxial strain is described by   

\begin{equation}
    \epsilon_{uni} = \sqrt{ \Biggl( \frac{\epsilon_{xx}-\epsilon_{yy}}{2} \Biggl)^2+\epsilon_{xy}^2} = \frac{\epsilon_{1} - \epsilon_{2}}{2}.
    \label{dev}
\end{equation}
 
Whenever strain stretches an atomic lattice along one direction, the lattice contracts along the transverse directions. The ratio of the contraction to the stretch is Poisson's ratio, which for Ga$_2$Se$_2$ is 0.23. When we perform the uniaxial strain gauge analysis reported in the main manuscript, we first numerically screen the computed strain at every point to identify those for which the ratio of $\epsilon_{1}$ to $\epsilon_{2}$ is less than 0.15. This threshold is below Poisson's ratio, which means that the strain at such a point is uniaxial. We compute the magnitude of the uniaxial strain at that point using Equation \ref{dev} and then plot the measured PL peak shift as a function of the uniaxial strain amplitude to determine the uniaxial strain gauge factor $\alpha_{exp}$.

%\begin{equation}
%    \Delta E \approx \beta\frac{\epsilon_{xx}+\epsilon_{yy}}{2}
%    -
%    \alpha\frac{|\epsilon_{xx}|-|\epsilon_{yy}|}{2}.
%    \label{eq_bandgap}
%\end{equation}

Finally, we use the results of Equations \ref{hyd} and \ref{dev} to compose the model for bandgap shift as a function of local biaxial and uniaxial strain
\begin{equation}
    \Delta E \approx \beta_{exp}\frac{\epsilon_1+\epsilon_2}{2}+\alpha_{exp}\frac{|\epsilon_1-\epsilon_2|}{2}
    \label{PL peak shift}
\end{equation}

\section{Normalized RMSE Calculation}
To assess the performance of our model we calculated the normalized root mean square error (RMSE) between the experimentally-measured peak shift and the peak shift predicted by Equation \ref{PL peak shift} using the local computed strain and the experimentally-determined strain gauge factors. The RMSE is calculated using

\begin{equation}
    RMSE=\sqrt{\frac{1}{N}\sum_i(\hat{y_i}-y_i)^2}
\end{equation}

where $\hat{y_i}$ is the peak shift predicted by Equation \ref{PL peak shift}, $y_i$ is the experimentally-measured peak shift, and $N$ is the number of points. Because the range of both predicted and experimentally-measured peak shifts vary significantly from structure to structure based on the amount of strain induced, we normalized the RMSE computed for each structure by dividing by the range of the peak shift for each structure. This normalized RMSE provides a means for comparing the accuracy of the model described by Equation \ref{PL peak shift} across all structures. 

\subsection{COMSOL Comparison to AFM}
To probe one potential source of error in our model, we looked at the difference between the experimentally determined height profile (measured via AFM) and the extracted height from the COMSOL simulations. We then computed the normalized RMSE values across all structures, which was 7.22\%. 

\begin{figure}[h]
  \includegraphics[width=1\linewidth]{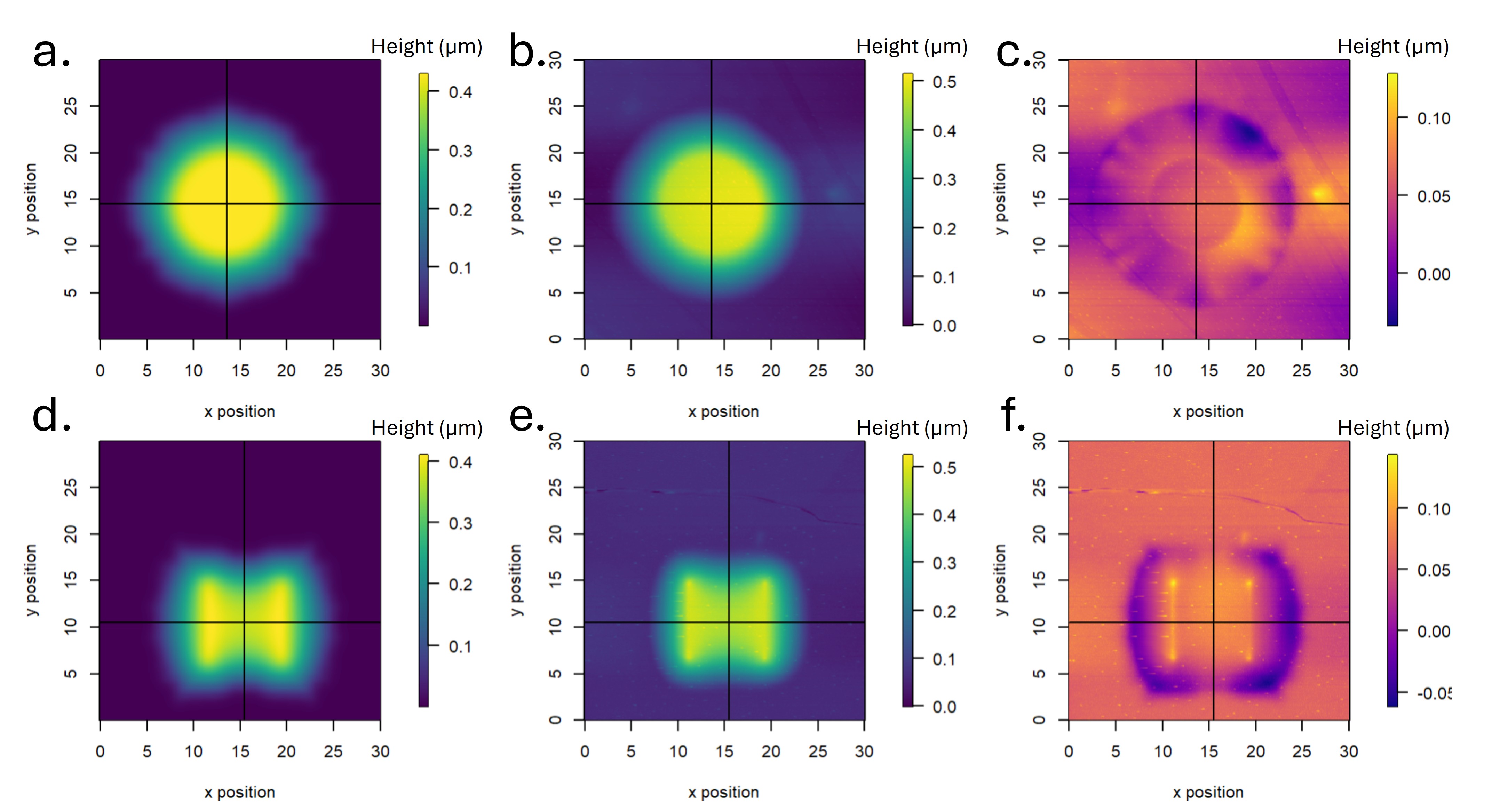}
  \caption{(a,d) The height profile extracted from the COMSOL simulations for an example ring and ridge structure. (b,e) The AFM height profile for the same example ring and ridge structures. (c,f) The difference between the experimental and predicted heights.}
  \label{afm_rmse}
\end{figure}

\section{All Structures: Experimental versus Model}
\begin{figure}[h]
  \includegraphics[width=1\linewidth]{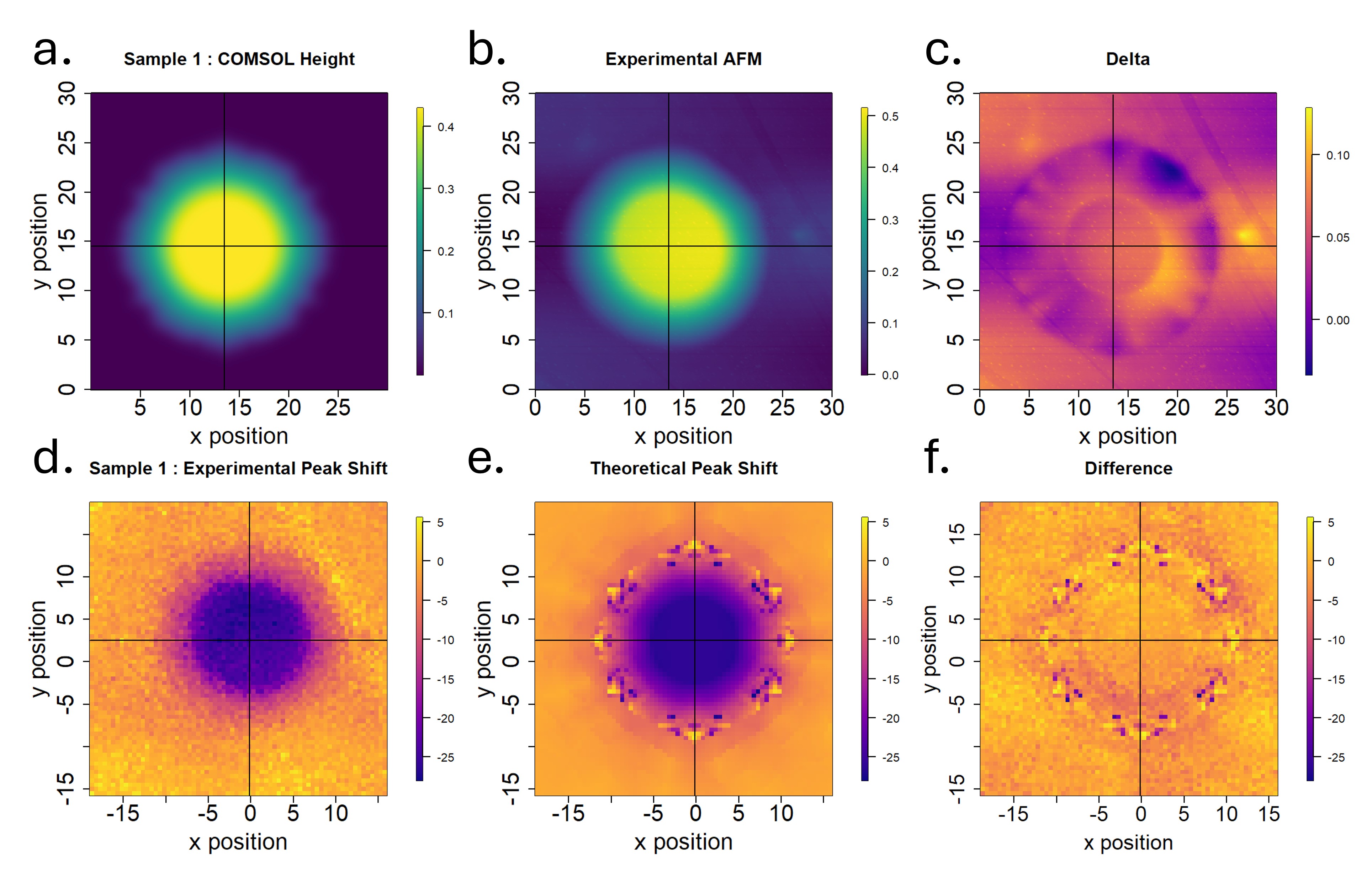}
  \caption{Sample 1. The height profile (a) simulated from COMSOL and (b) measured  by AFM, with (c) showing the difference between the experimental and predicted heights. (d) The experimental peak shift, (e) the calculated peak shift using the experimentally determined strain gauge factors, and (f) the difference between the two.}
  \label{sample_01}
\end{figure}

\begin{figure}[h]
  \includegraphics[width=1\linewidth]{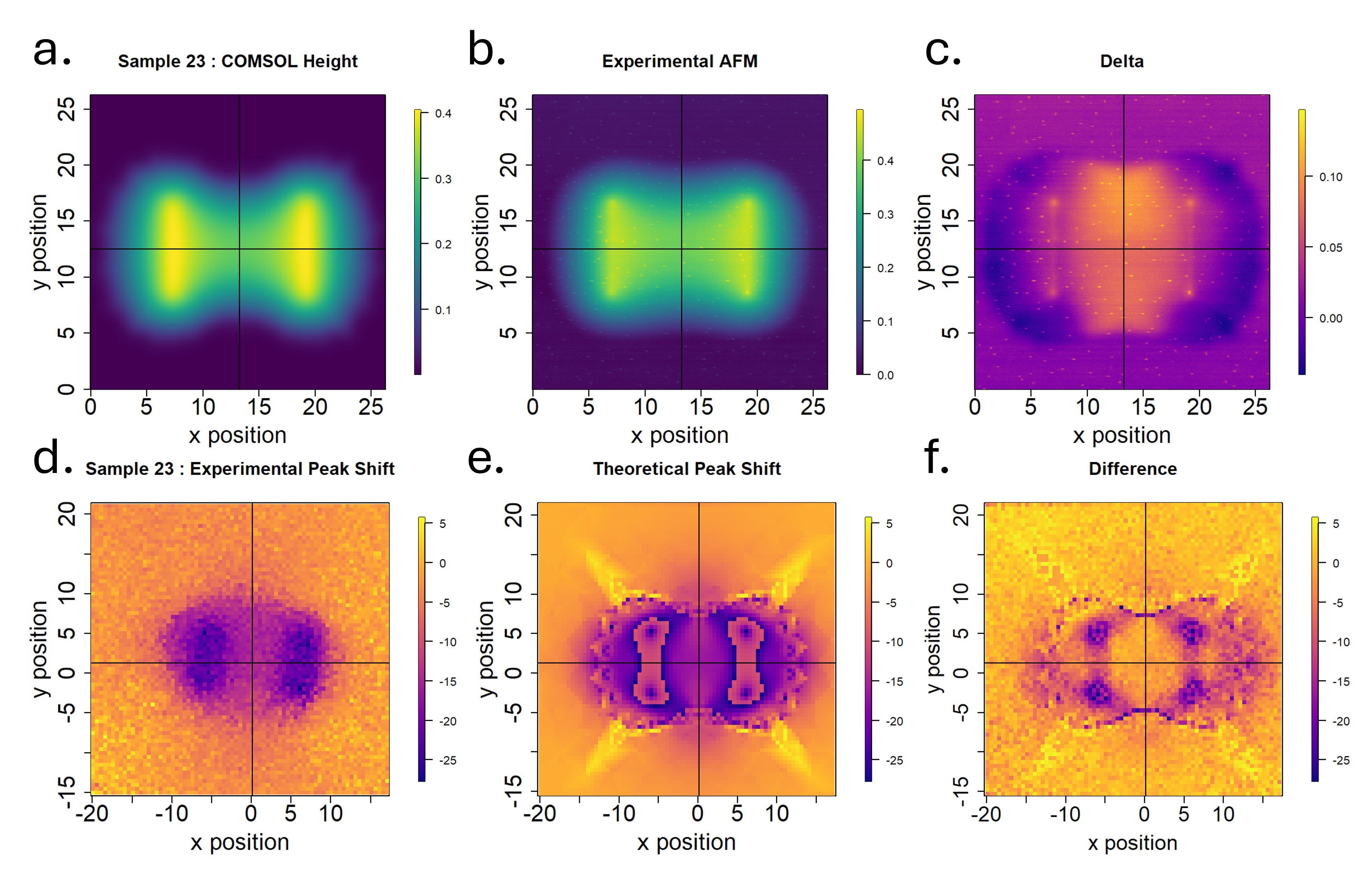}
  \caption{Sample 23. The height profile (a) simulated from COMSOL and (b) measured  by AFM, with (c) showing the difference between the experimental and predicted heights. (d) The experimental peak shift, (e) the calculated peak shift using the experimentally determined strain gauge factors, and (f) the difference between the two.}
  \label{sample_23}
\end{figure}

In Figure S\ref{sample_01}-S\ref{sample_23} we provide two-dimensional plots of the experimental and simulated results for each sample measured. In each case panels (a-c) show the height profiles from (a) the COMSOL simulations and (b) the AFM scan, with (c) showing the difference between the two. In each case panels (d-f) show (d) the experimentally measured PL peak shift, (e) the theoretical PL peak shift computed using Equation \ref{PL peak shift} and the experimentally determined strain gauge factors, and (f) the difference between the two. The main differences occur at points of high strain (e.g.~tips of the ridges) or where we see numerical artifacts in the COMSOL simulations (e.g.~where the flake touches the substrate).

\begin{figure}[h]
  \includegraphics[width=1\linewidth]{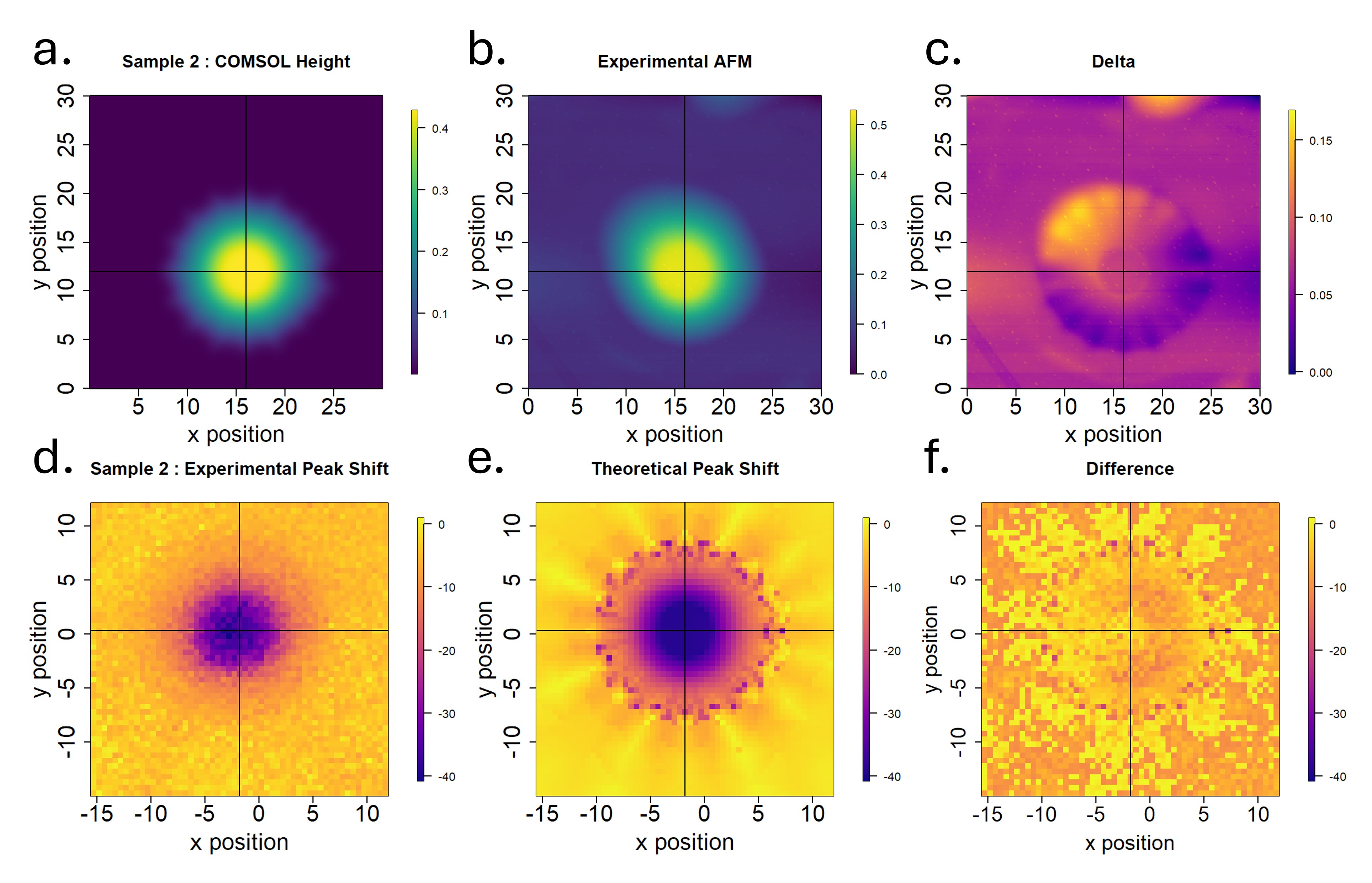}
  \caption{Sample 2. The height profile (a) simulated from COMSOL and (b) measured  by AFM, with (c) showing the difference between the experimental and predicted heights. (d) The experimental peak shift, (e) the calculated peak shift using the experimentally determined strain gauge factors, and (f) the difference between the two.}
  \label{sample_02}
\end{figure}
\begin{figure}[h]
  \includegraphics[width=1\linewidth]{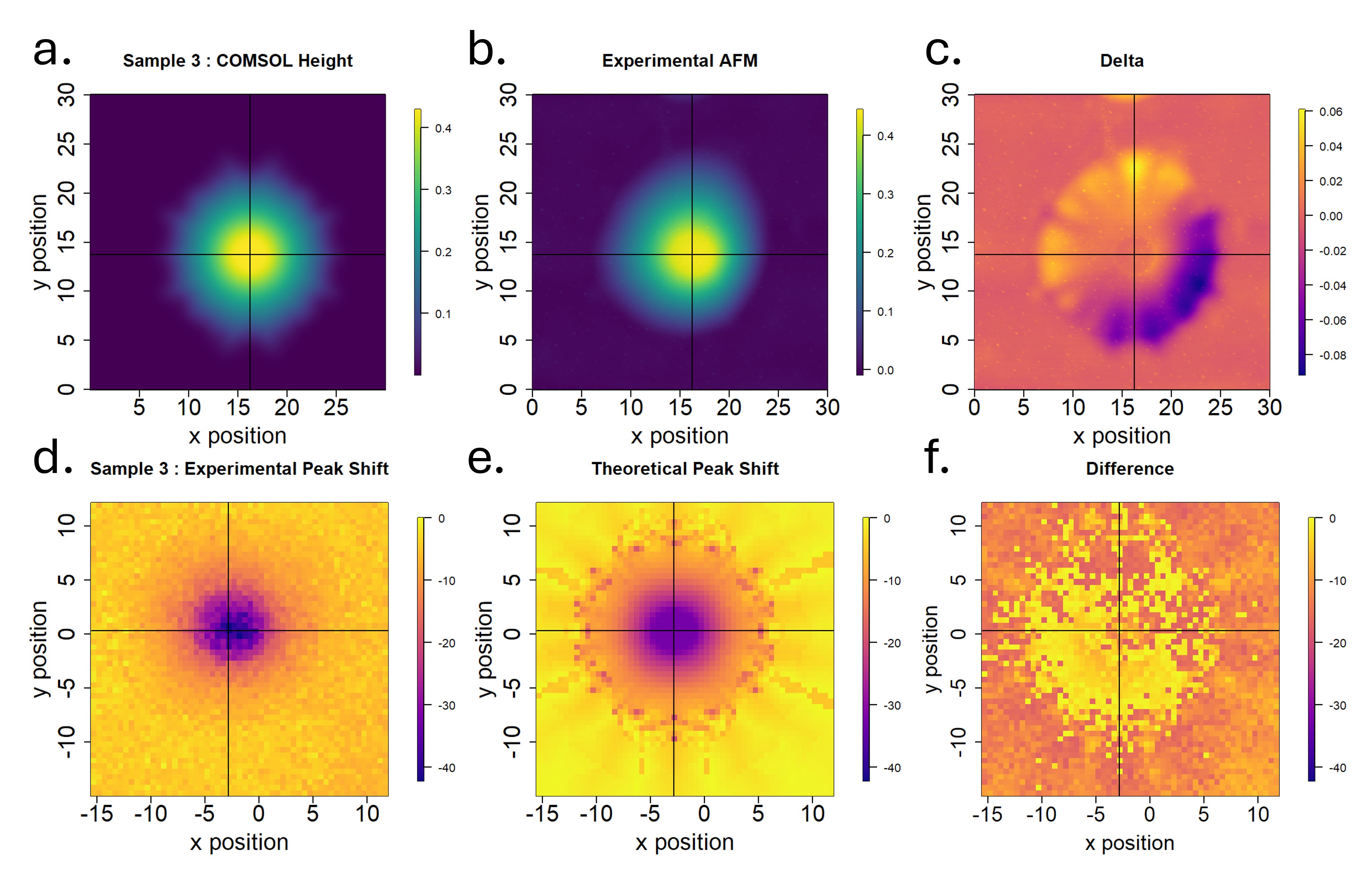}
  \caption{Sample 3. The height profile (a) simulated from COMSOL and (b) measured  by AFM, with (c) showing the difference between the experimental and predicted heights. (d) The experimental peak shift, (e) the calculated peak shift using the experimentally determined strain gauge factors, and (f) the difference between the two.}
  \label{sample_03}
\end{figure}
\begin{figure}[h]
  \includegraphics[width=1\linewidth]{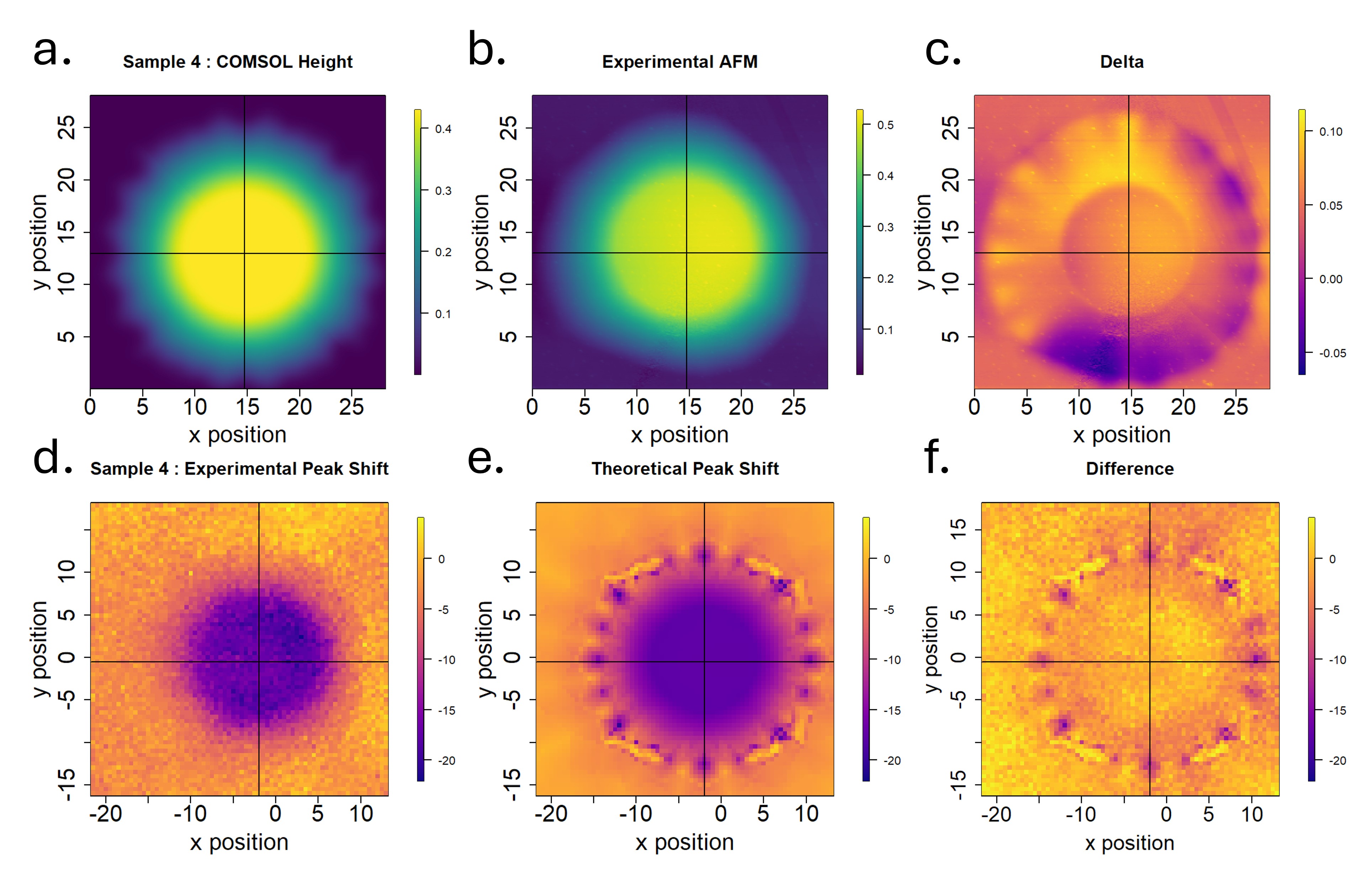}
  \caption{Sample 4. The height profile (a) simulated from COMSOL and (b) measured  by AFM, with (c) showing the difference between the experimental and predicted heights. (d) The experimental peak shift, (e) the calculated peak shift using the experimentally determined strain gauge factors, and (f) the difference between the two.}
  \label{sample_04}
\end{figure}
\begin{figure}[h]
  \includegraphics[width=1\linewidth]{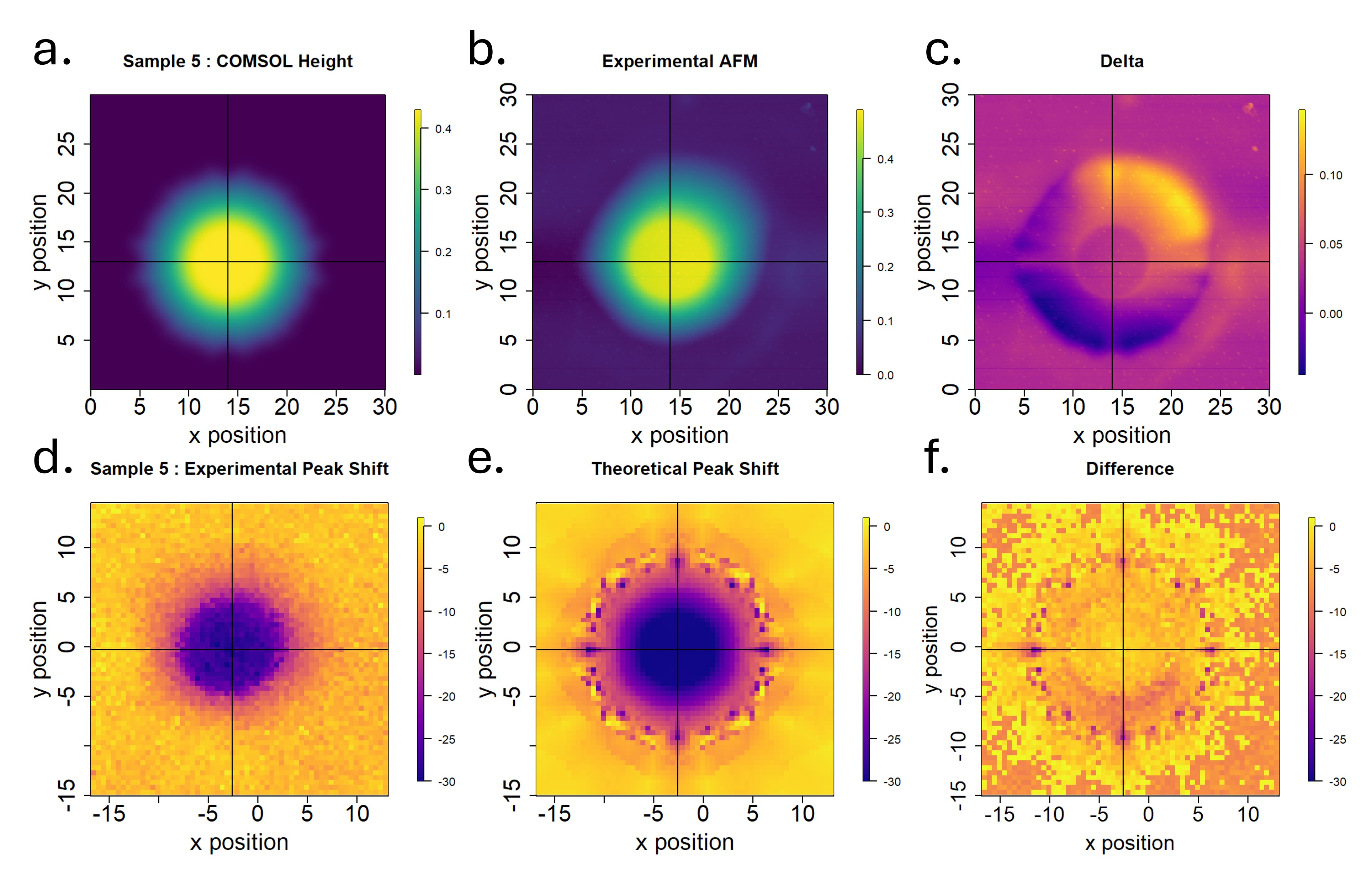}
  \caption{Sample 5. The height profile (a) simulated from COMSOL and (b) measured  by AFM, with (c) showing the difference between the experimental and predicted heights. (d) The experimental peak shift, (e) the calculated peak shift using the experimentally determined strain gauge factors, and (f) the difference between the two.}
  \label{sample_05}
\end{figure}
\begin{figure}[h]
  \includegraphics[width=1\linewidth]{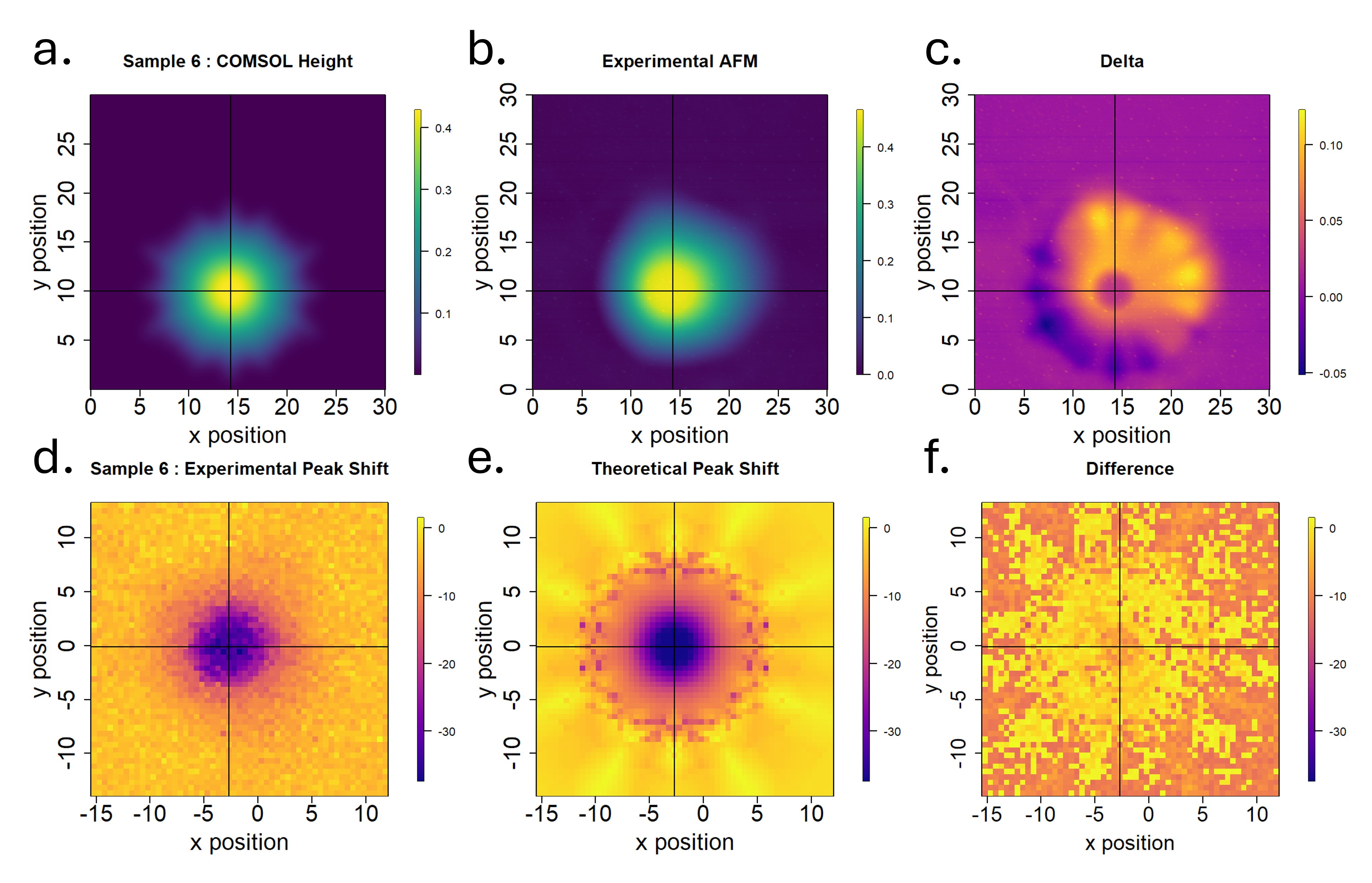}
  \caption{Sample 6. The height profile (a) simulated from COMSOL and (b) measured  by AFM, with (c) showing the difference between the experimental and predicted heights. (d) The experimental peak shift, (e) the calculated peak shift using the experimentally determined strain gauge factors, and (f) the difference between the two.}
  \label{sample_06}
\end{figure}
\begin{figure}[h]
  \includegraphics[width=1\linewidth]{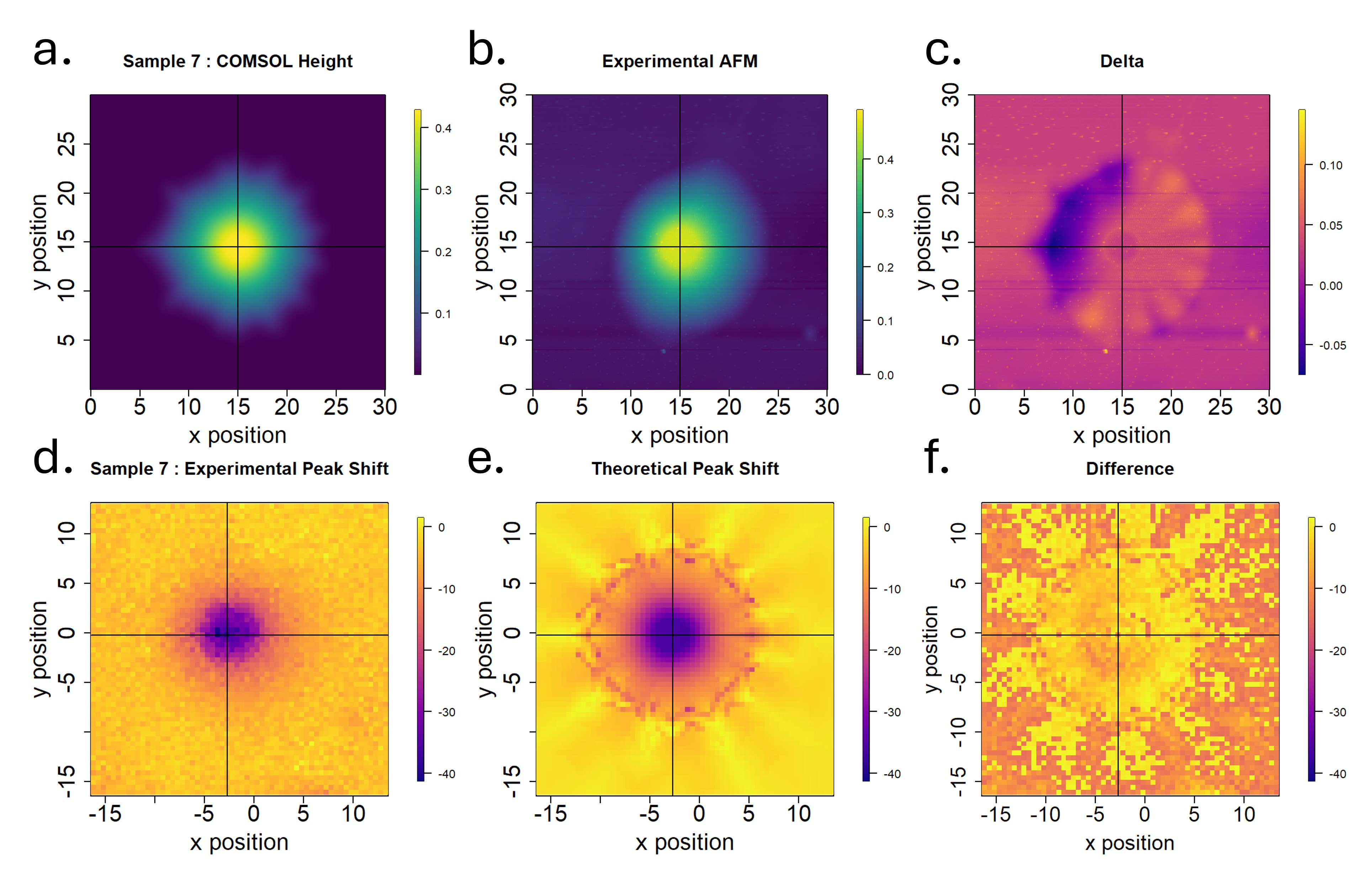}
  \caption{Sample 7. The height profile (a) simulated from COMSOL and (b) measured  by AFM, with (c) showing the difference between the experimental and predicted heights. (d) The experimental peak shift, (e) the calculated peak shift using the experimentally determined strain gauge factors, and (f) the difference between the two.}
  \label{sample_07}
\end{figure}
\begin{figure}[h]
  \includegraphics[width=1\linewidth]{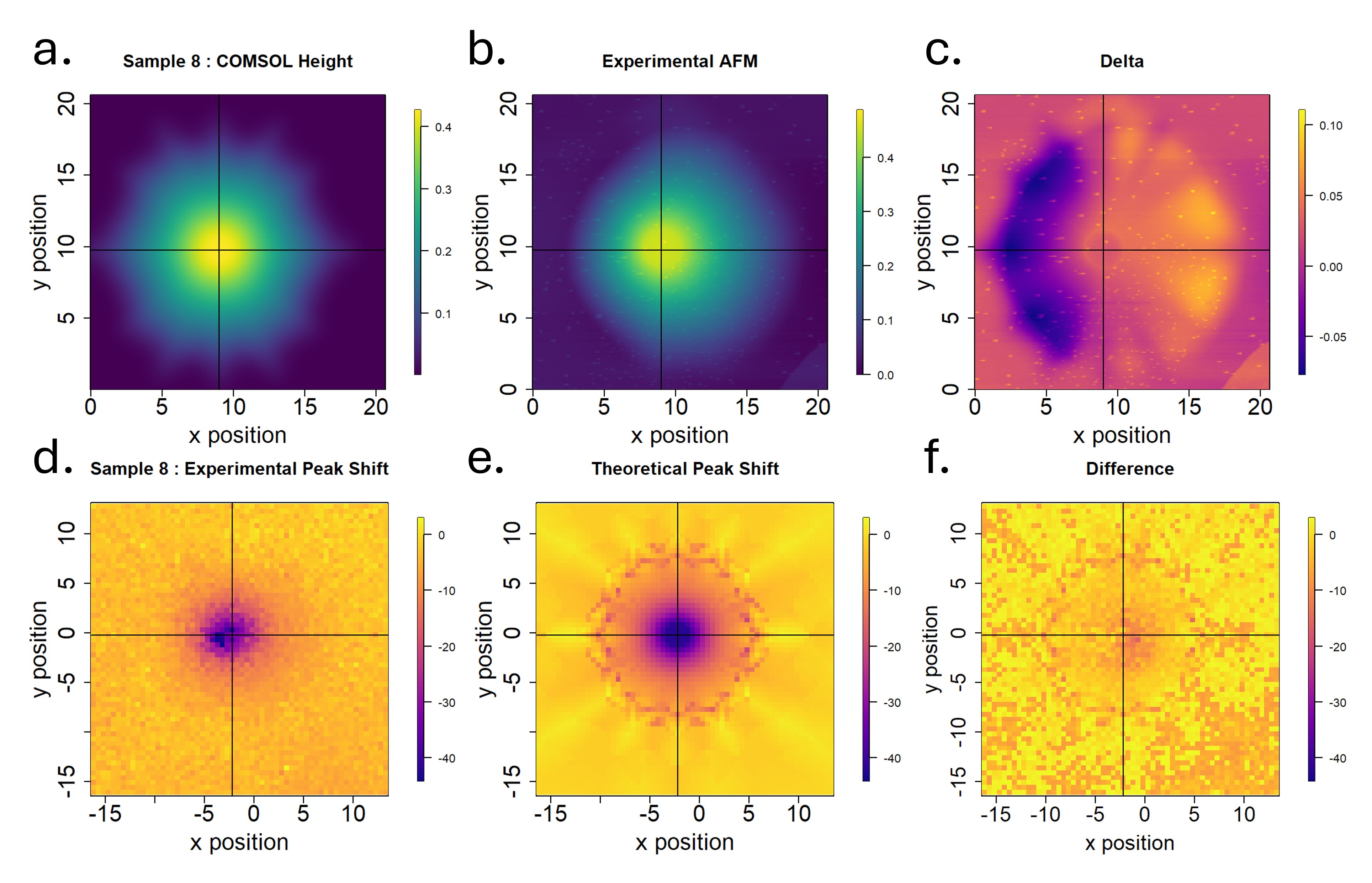}
  \caption{Sample 8. The height profile (a) simulated from COMSOL and (b) measured  by AFM, with (c) showing the difference between the experimental and predicted heights. (d) The experimental peak shift, (e) the calculated peak shift using the experimentally determined strain gauge factors, and (f) the difference between the two.}
  \label{sample_08}
\end{figure}
\begin{figure}[h]
  \includegraphics[width=1\linewidth]{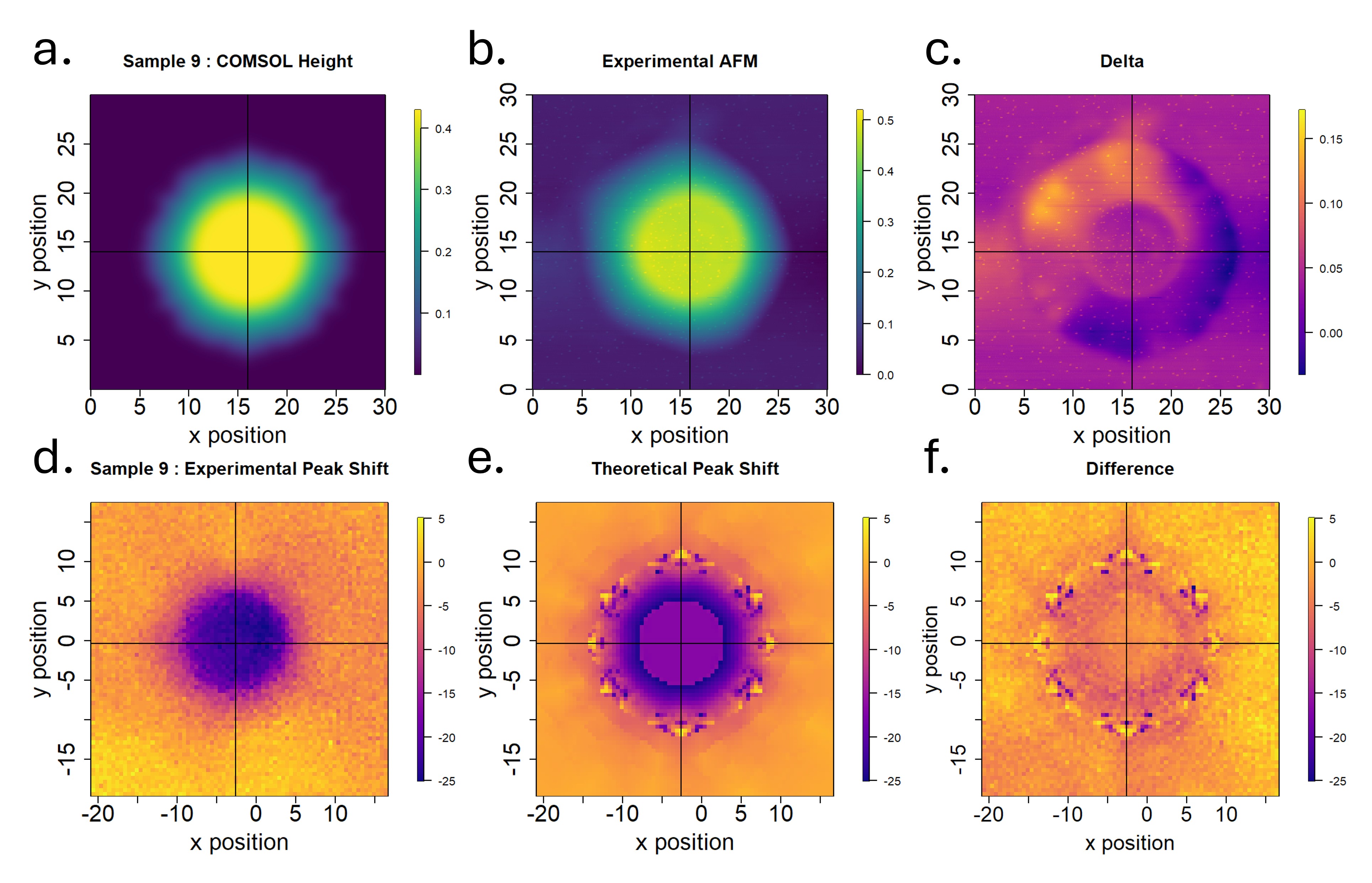}
  \caption{Sample 9. The height profile (a) simulated from COMSOL and (b) measured  by AFM, with (c) showing the difference between the experimental and predicted heights. (d) The experimental peak shift, (e) the calculated peak shift using the experimentally determined strain gauge factors, and (f) the difference between the two.}
  \label{sample_09}
\end{figure}
\begin{figure}[h]
  \includegraphics[width=1\linewidth]{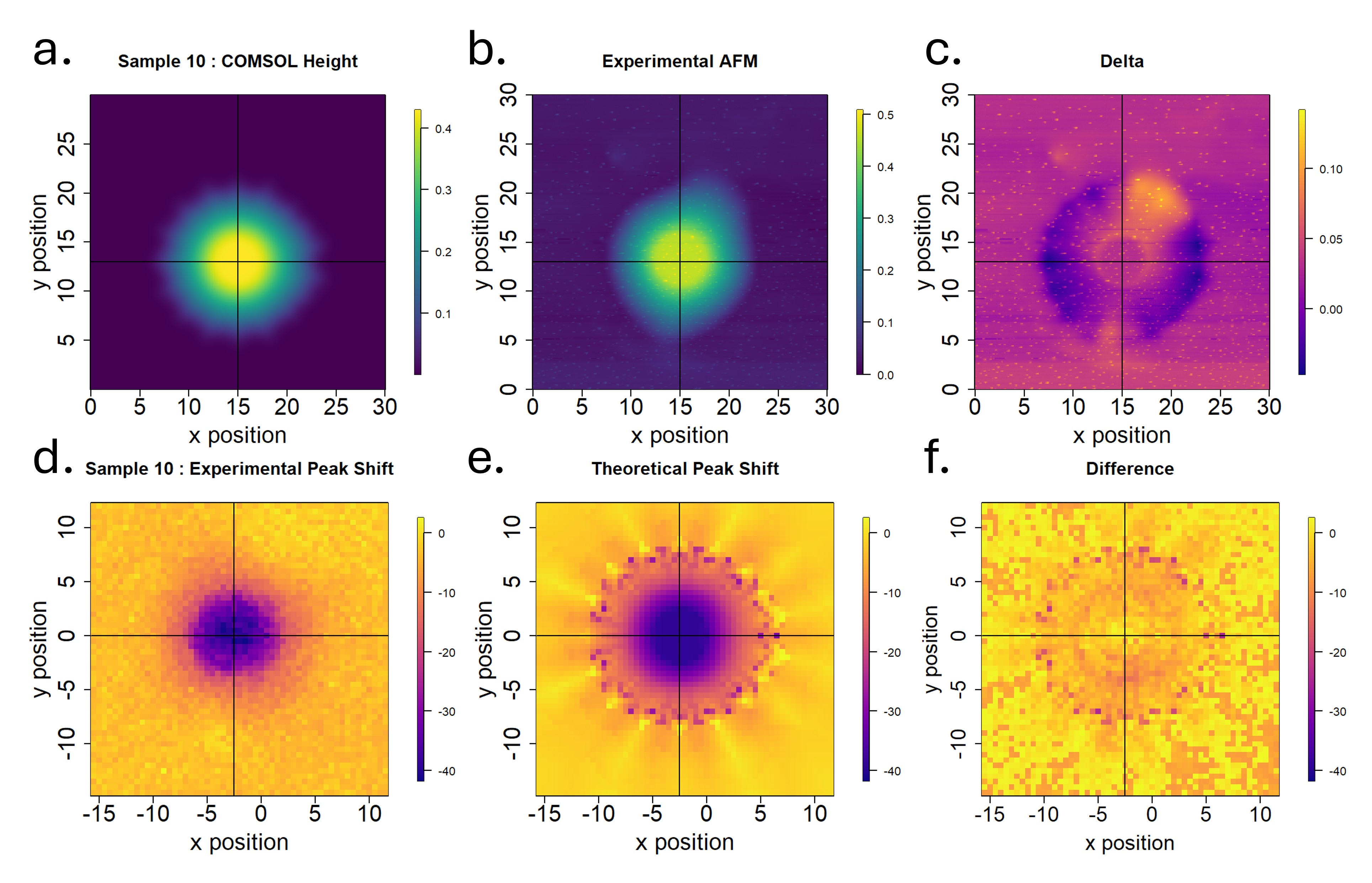}
  \caption{Sample 10. The height profile (a) simulated from COMSOL and (b) measured  by AFM, with (c) showing the difference between the experimental and predicted heights. (d) The experimental peak shift, (e) the calculated peak shift using the experimentally determined strain gauge factors, and (f) the difference between the two.}
  \label{sample_10}
\end{figure}

\begin{figure}[h]
  \includegraphics[width=1\linewidth]{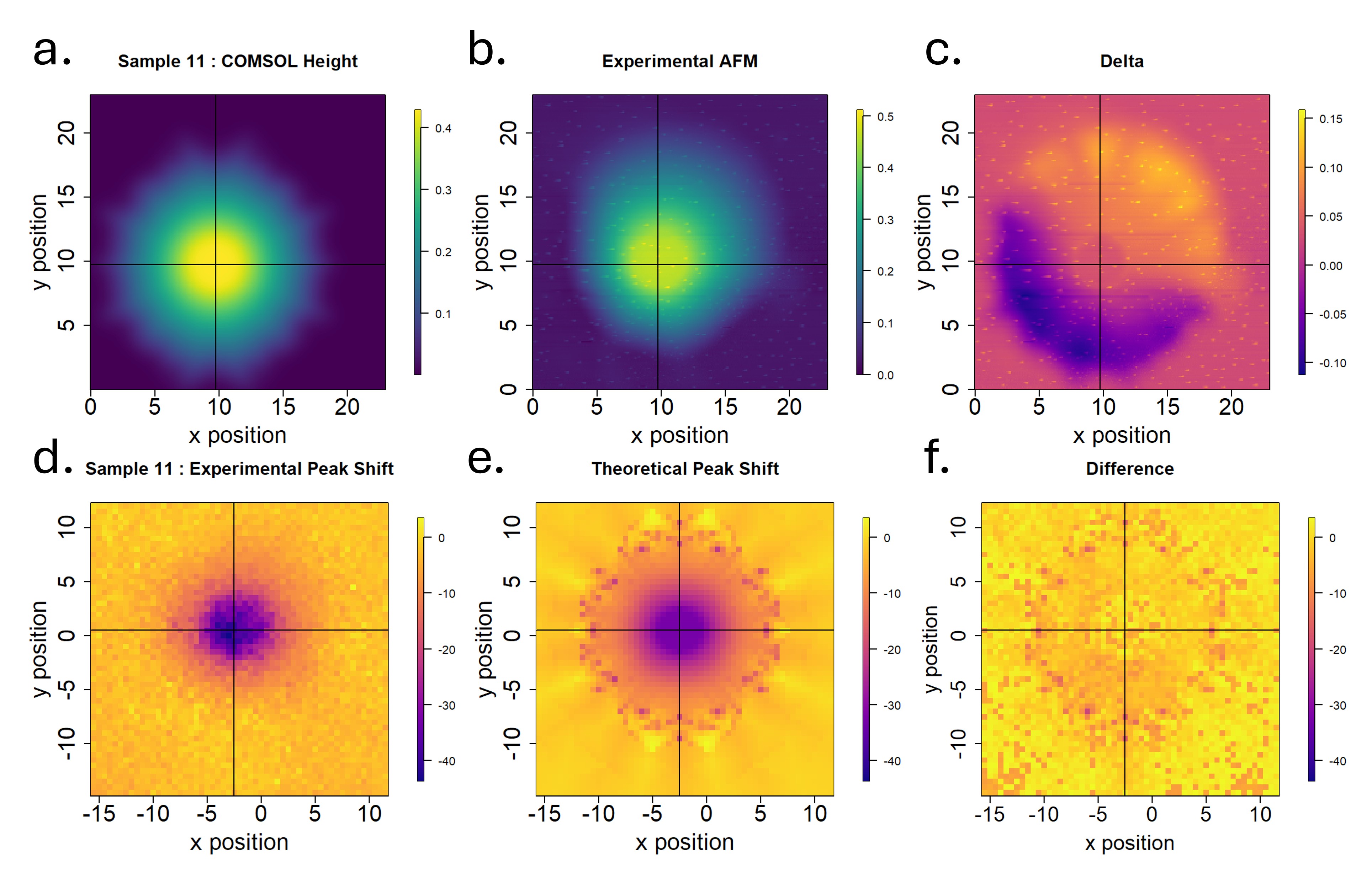}
  \caption{Sample 11. The height profile (a) simulated from COMSOL and (b) measured  by AFM, with (c) showing the difference between the experimental and predicted heights. (d) The experimental peak shift, (e) the calculated peak shift using the experimentally determined strain gauge factors, and (f) the difference between the two.}
  \label{sample_11}
\end{figure}
\begin{figure}[h]
  \includegraphics[width=1\linewidth]{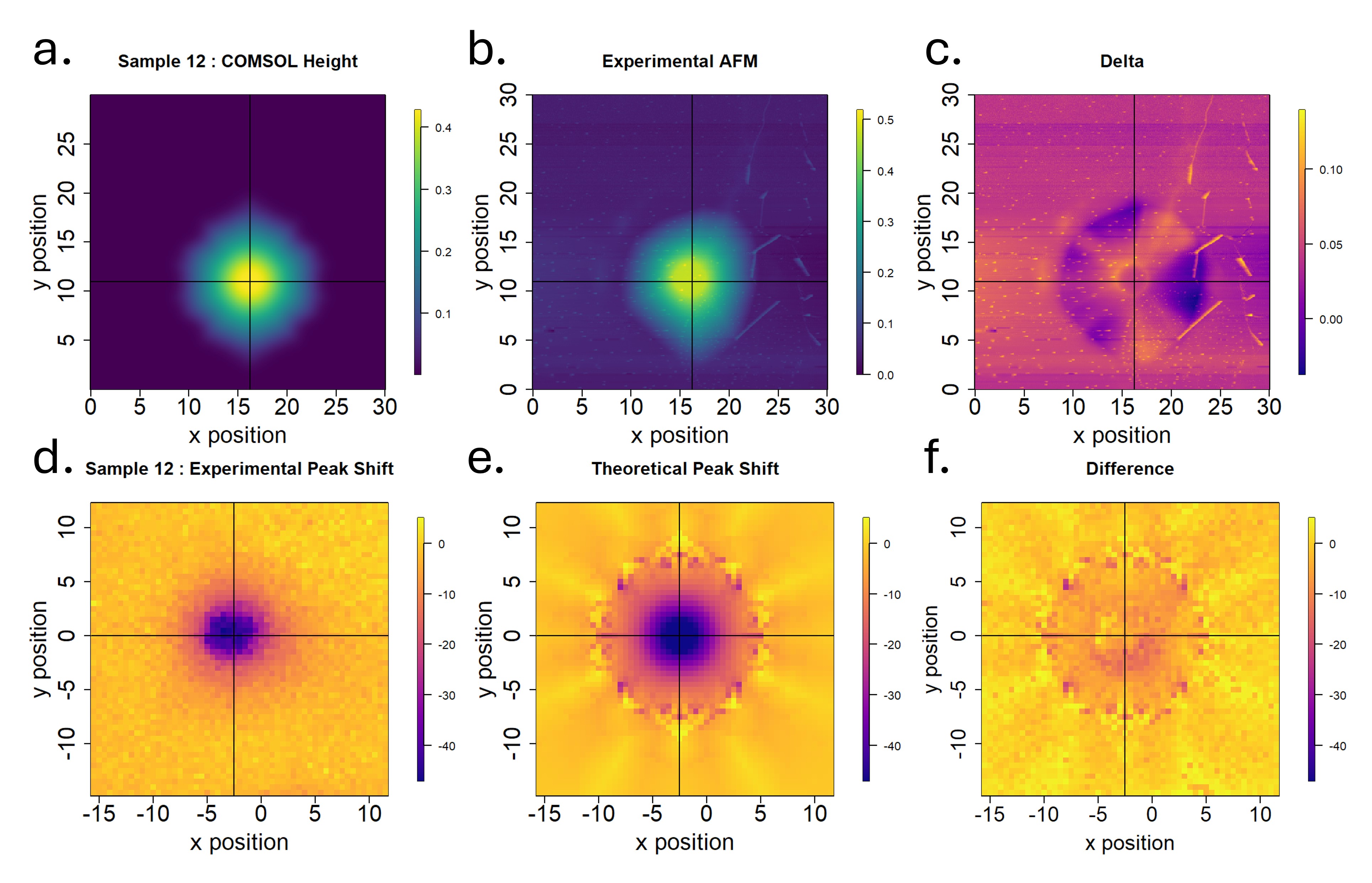}
  \caption{Sample 12. The height profile (a) simulated from COMSOL and (b) measured  by AFM, with (c) showing the difference between the experimental and predicted heights. (d) The experimental peak shift, (e) the calculated peak shift using the experimentally determined strain gauge factors, and (f) the difference between the two.}
  \label{sample_12}
\end{figure}
\begin{figure}[h]
  \includegraphics[width=1\linewidth]{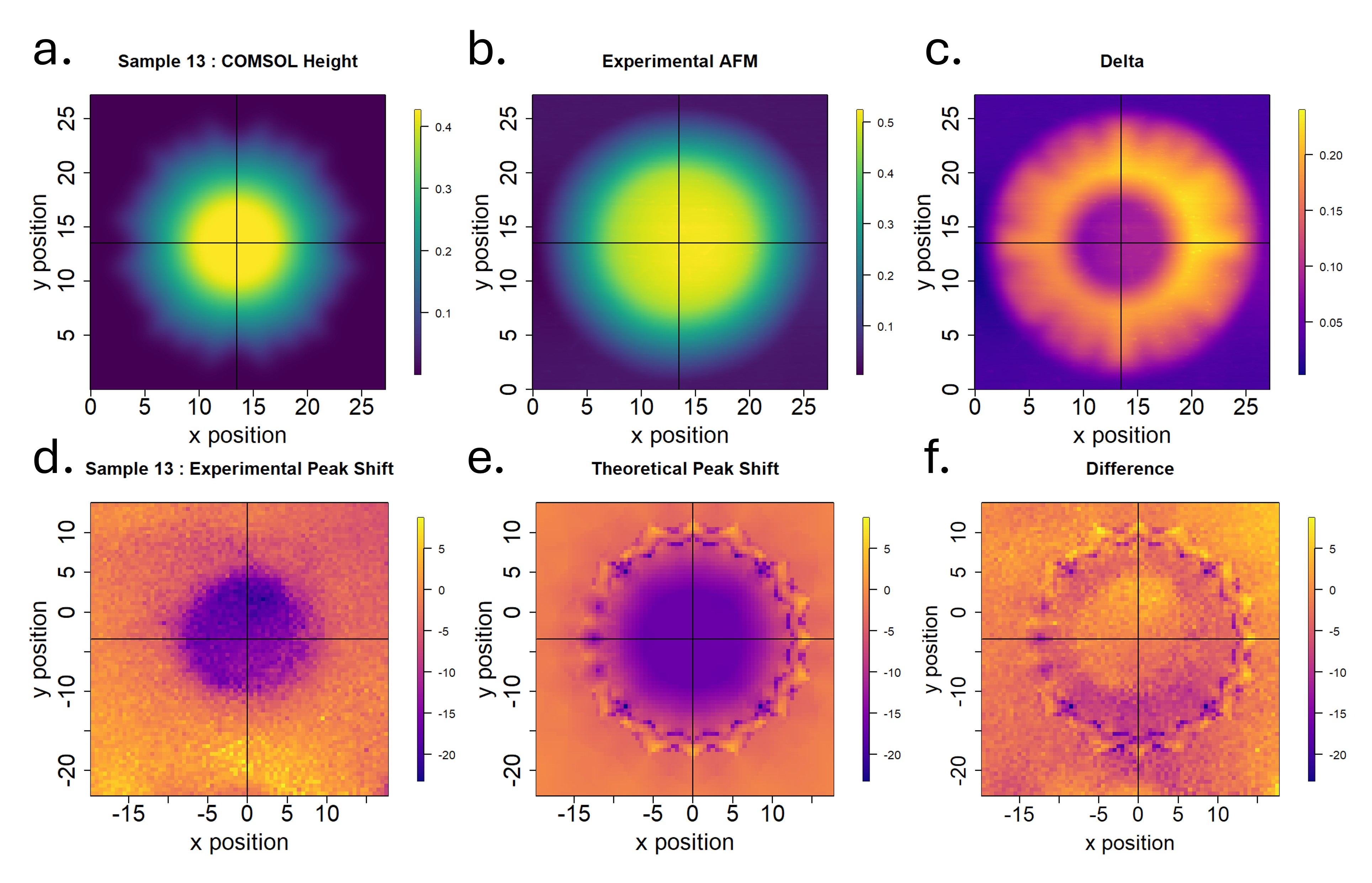}
  \caption{Sample 13. The height profile (a) simulated from COMSOL and (b) measured  by AFM, with (c) showing the difference between the experimental and predicted heights. (d) The experimental peak shift, (e) the calculated peak shift using the experimentally determined strain gauge factors, and (f) the difference between the two.}
  \label{sample_13}
\end{figure}
\begin{figure}[h]
  \includegraphics[width=1\linewidth]{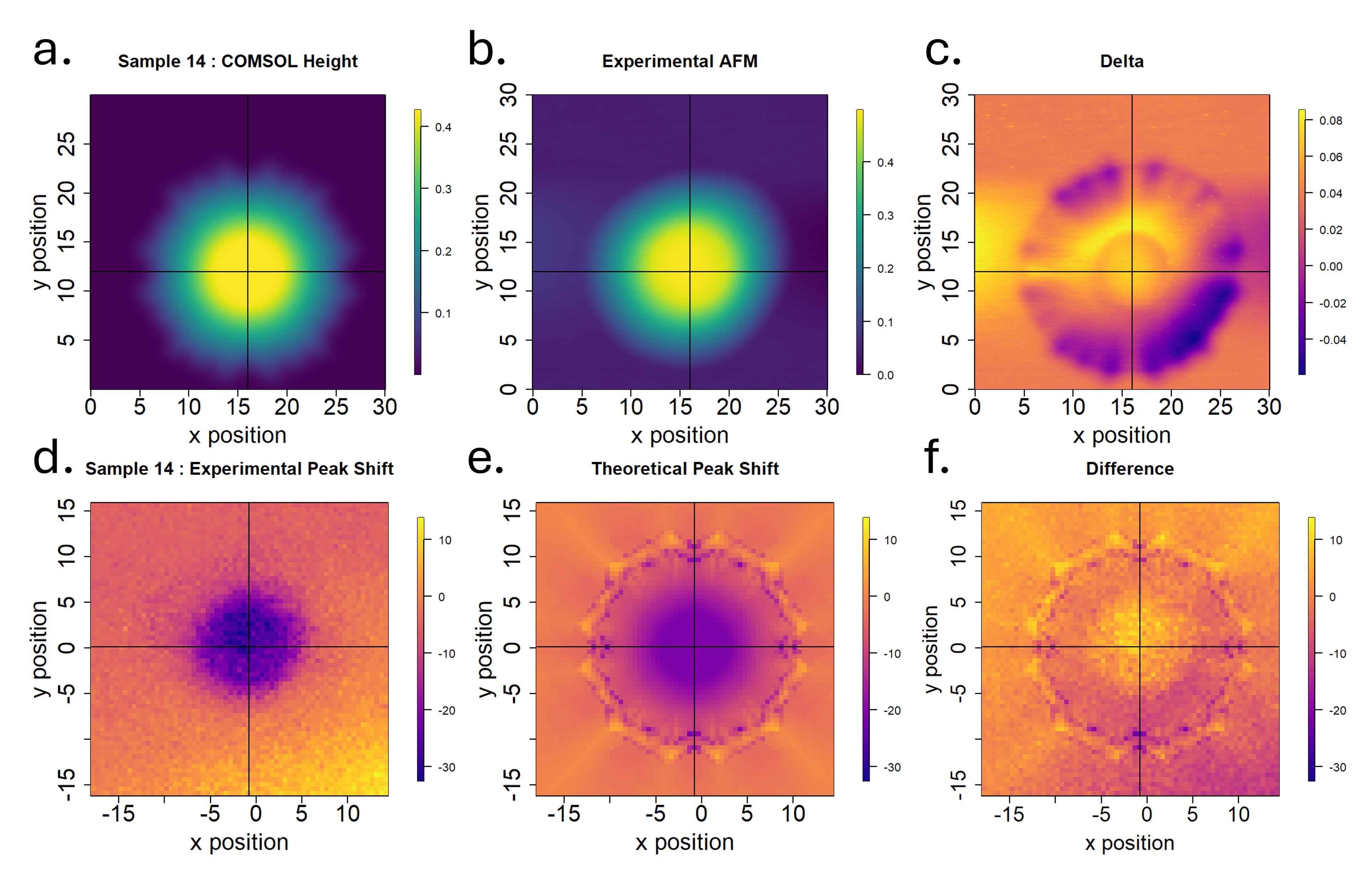}
  \caption{Sample 14. The height profile (a) simulated from COMSOL and (b) measured  by AFM, with (c) showing the difference between the experimental and predicted heights. (d) The experimental peak shift, (e) the calculated peak shift using the experimentally determined strain gauge factors, and (f) the difference between the two.}
  \label{sample_14}
\end{figure}
\begin{figure}[h]
  \includegraphics[width=1\linewidth]{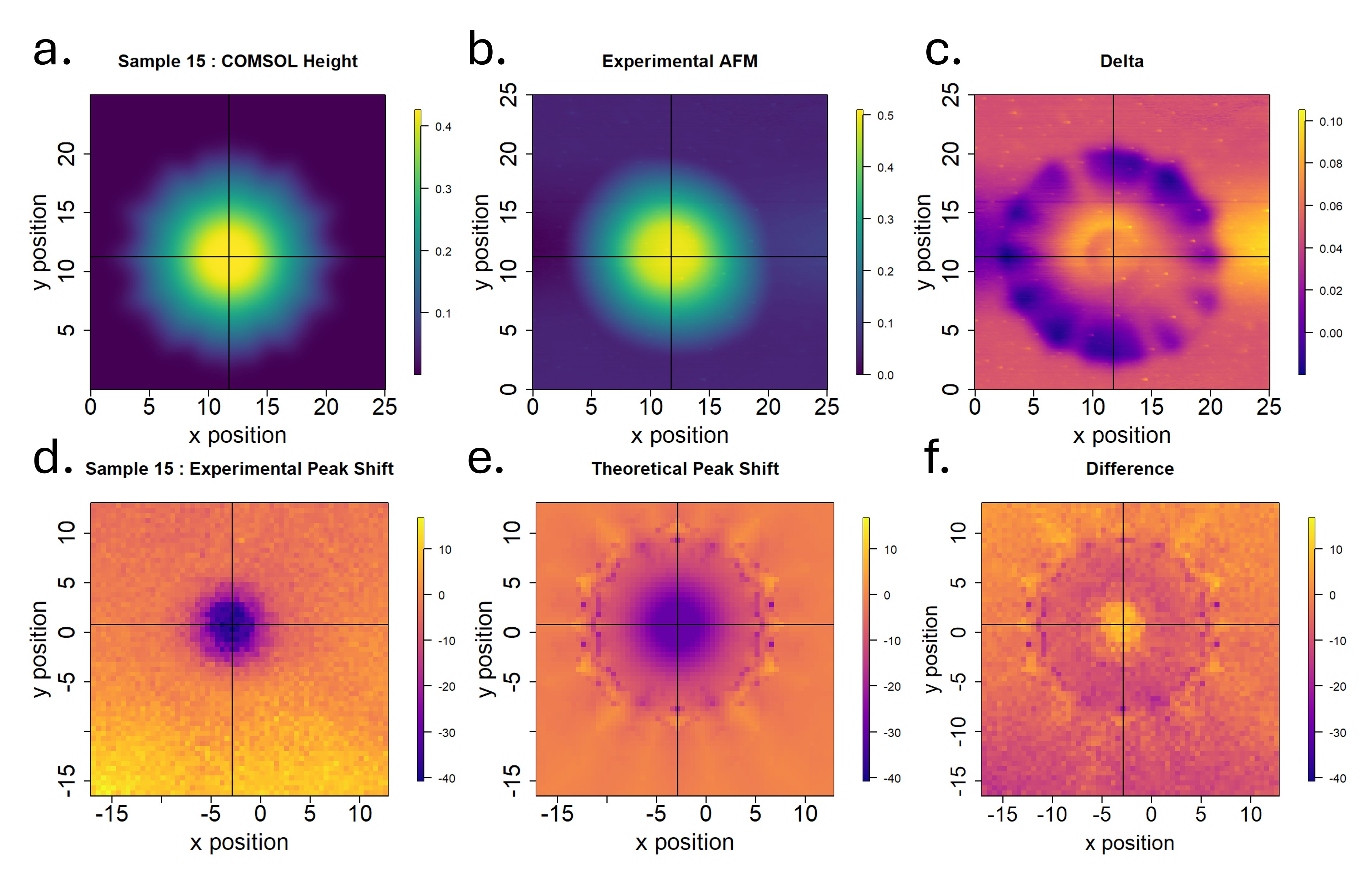}
  \caption{Sample 15. The height profile (a) simulated from COMSOL and (b) measured  by AFM, with (c) showing the difference between the experimental and predicted heights. (d) The experimental peak shift, (e) the calculated peak shift using the experimentally determined strain gauge factors, and (f) the difference between the two.}
  \label{sample_15}
\end{figure}
\begin{figure}[h]
  \includegraphics[width=1\linewidth]{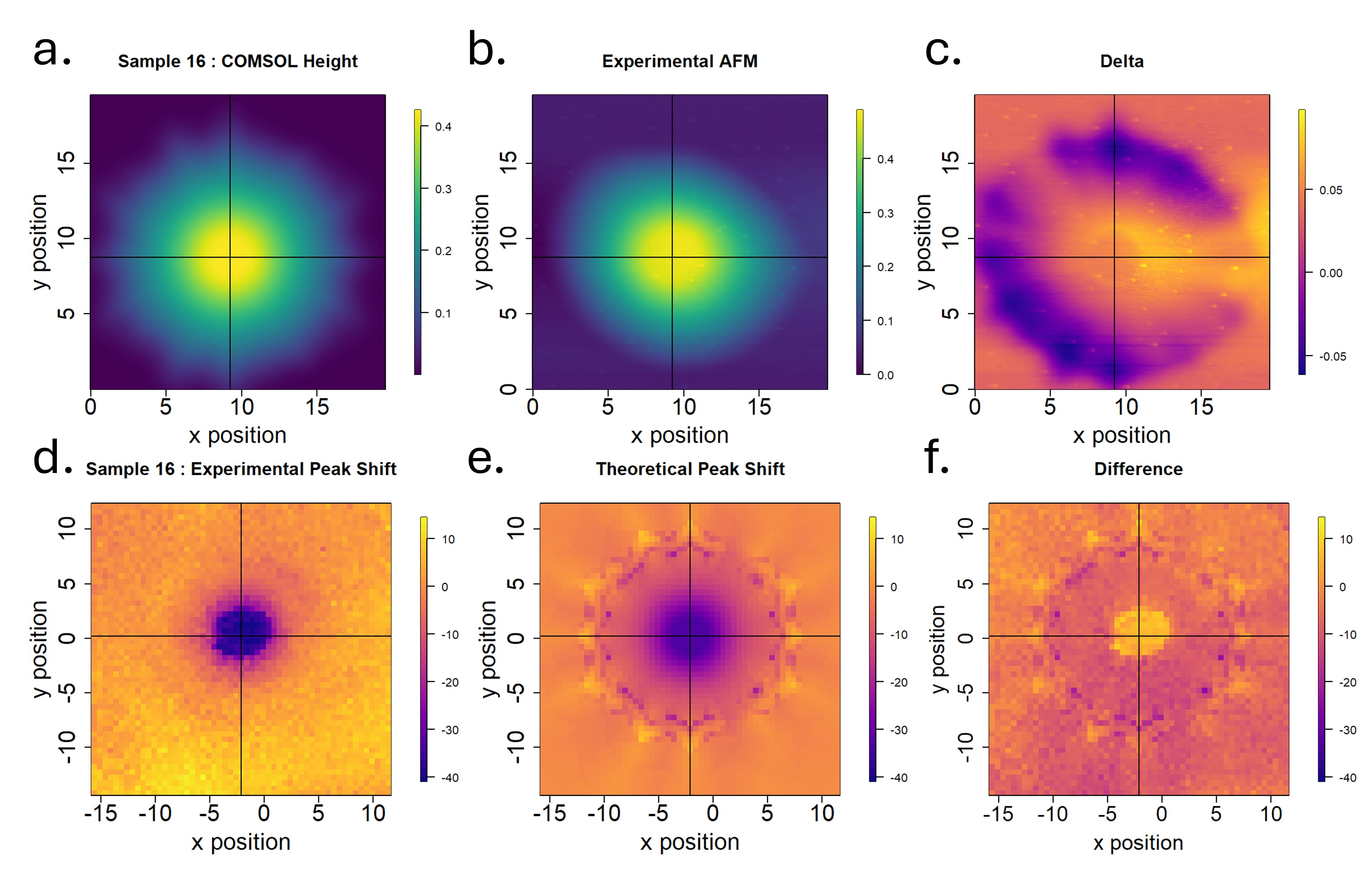}
  \caption{Sample 16. The height profile (a) simulated from COMSOL and (b) measured  by AFM, with (c) showing the difference between the experimental and predicted heights. (d) The experimental peak shift, (e) the calculated peak shift using the experimentally determined strain gauge factors, and (f) the difference between the two.}
  \label{sample_16}
\end{figure}
\begin{figure}[h]
  \includegraphics[width=1\linewidth]{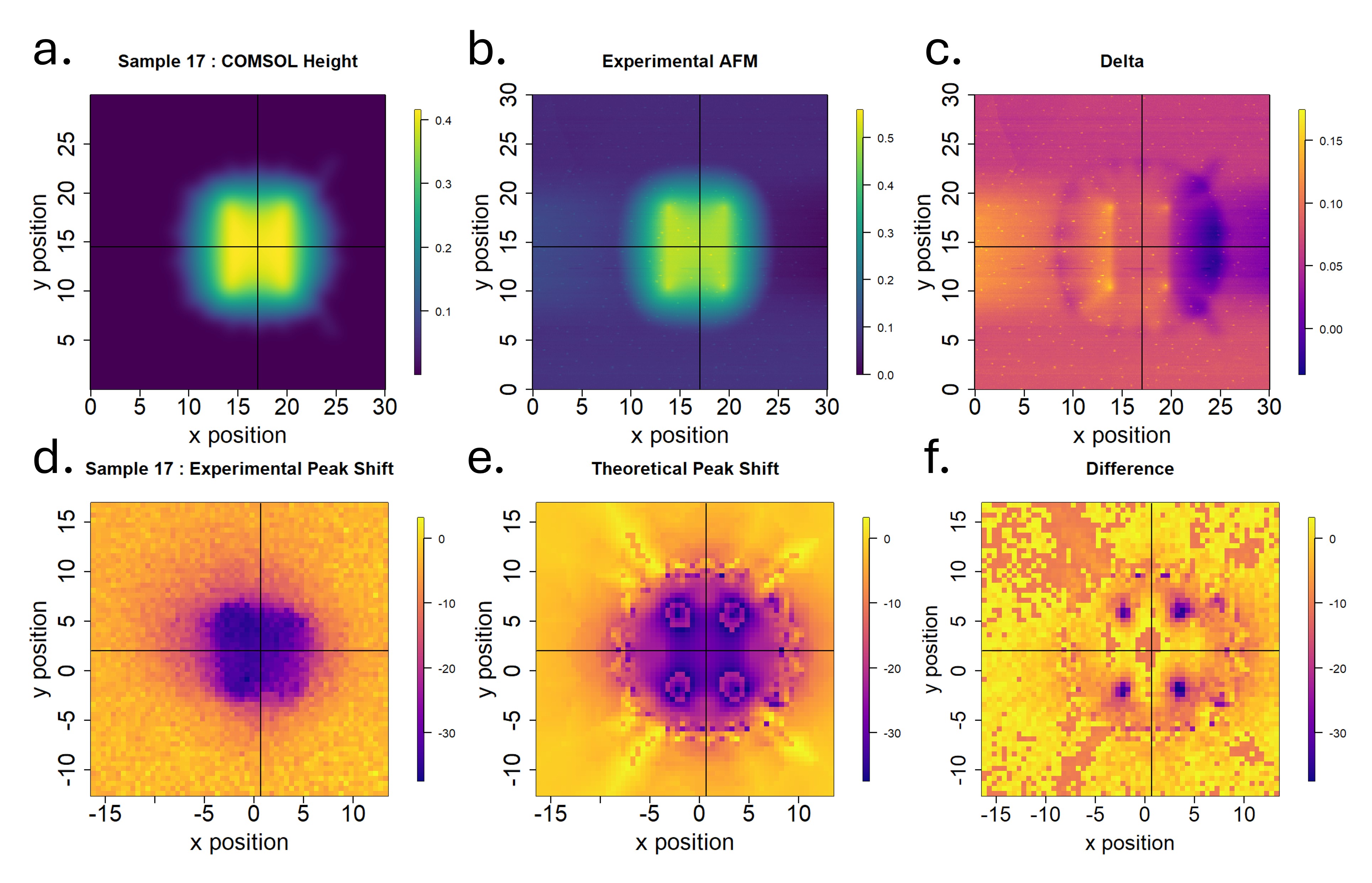}
  \caption{Sample 17. The height profile (a) simulated from COMSOL and (b) measured  by AFM, with (c) showing the difference between the experimental and predicted heights. (d) The experimental peak shift, (e) the calculated peak shift using the experimentally determined strain gauge factors, and (f) the difference between the two.}
  \label{sample_17}
\end{figure}
\begin{figure}[h]
  \includegraphics[width=1\linewidth]{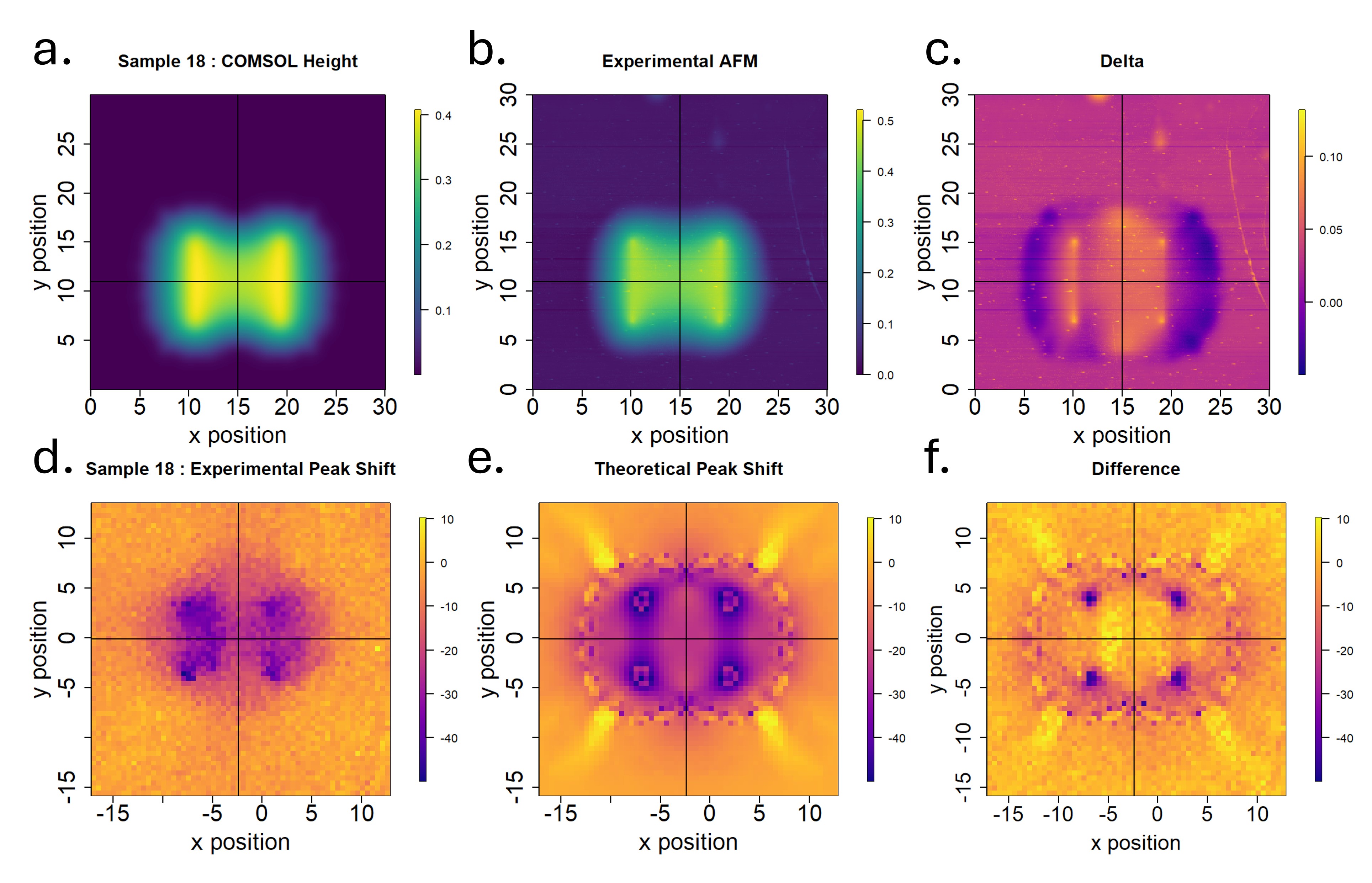}
  \caption{Sample 18. The height profile (a) simulated from COMSOL and (b) measured  by AFM, with (c) showing the difference between the experimental and predicted heights. (d) The experimental peak shift, (e) the calculated peak shift using the experimentally determined strain gauge factors, and (f) the difference between the two.}
  \label{sample_18}
\end{figure}
\begin{figure}[h]
  \includegraphics[width=1\linewidth]{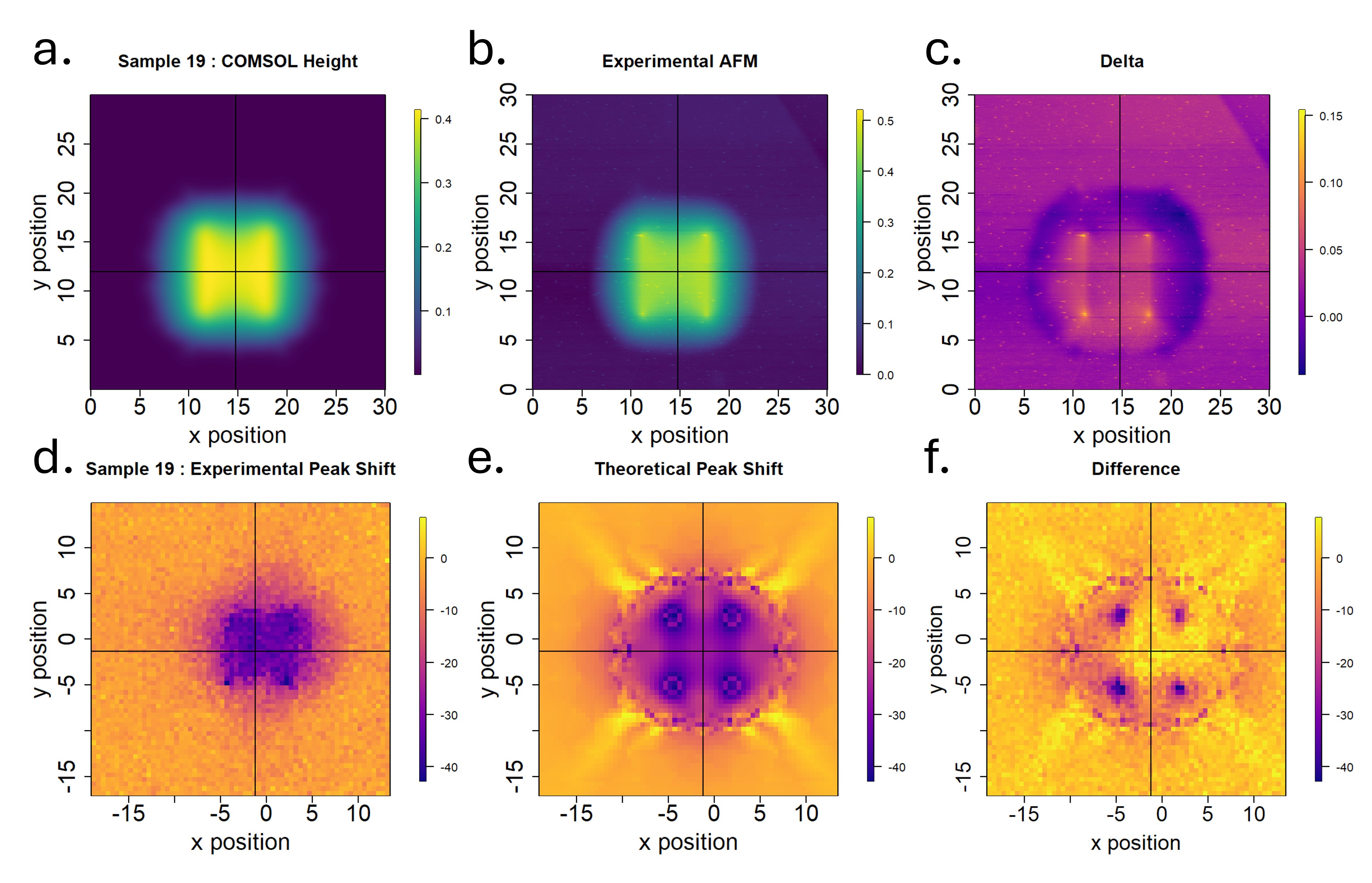}
  \caption{Sample 19. The height profile (a) simulated from COMSOL and (b) measured  by AFM, with (c) showing the difference between the experimental and predicted heights. (d) The experimental peak shift, (e) the calculated peak shift using the experimentally determined strain gauge factors, and (f) the difference between the two.}
  \label{sample_19}
\end{figure}
\begin{figure}[h]
  \includegraphics[width=1\linewidth]{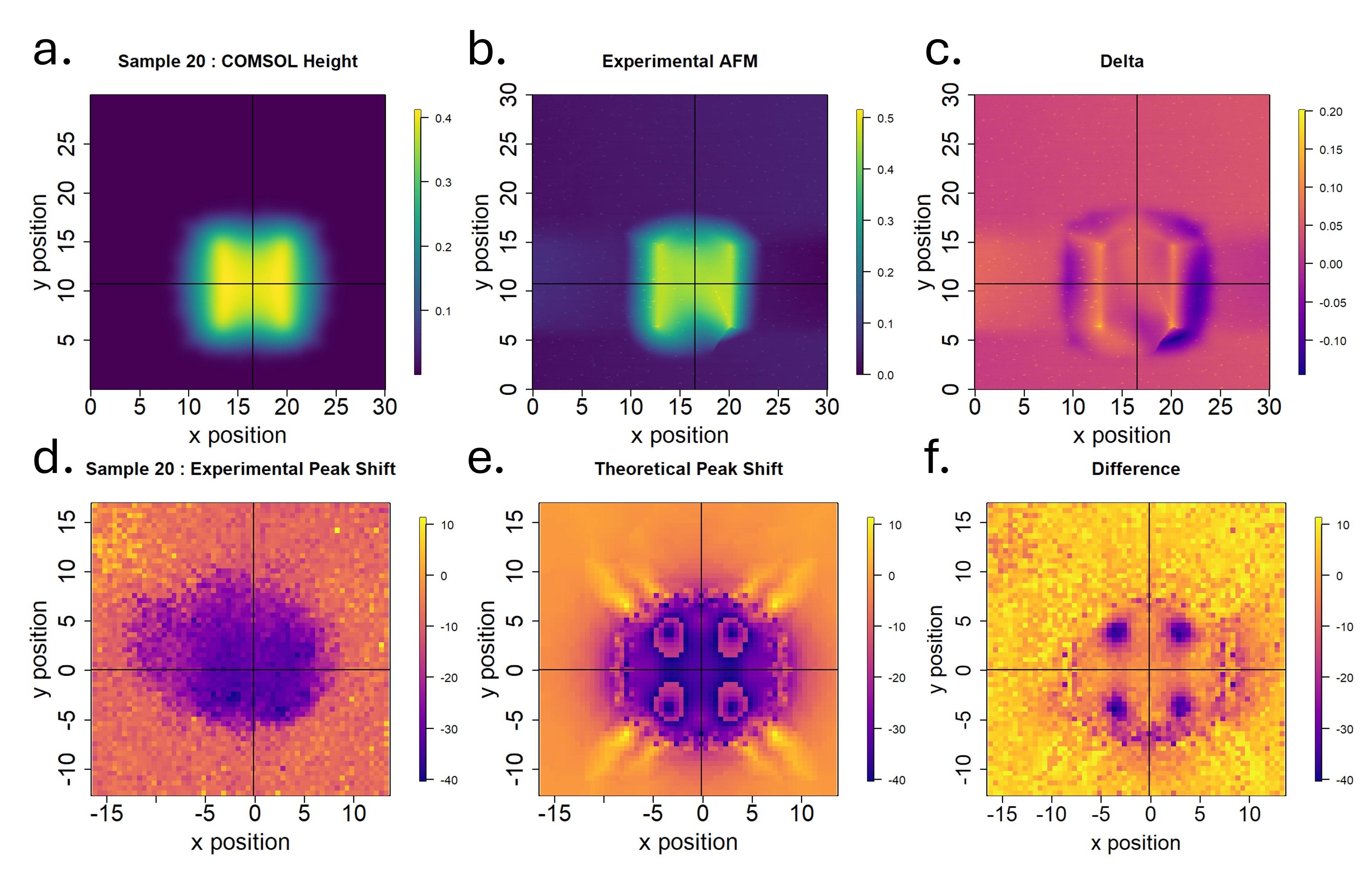}
  \caption{Sample 20. The height profile (a) simulated from COMSOL and (b) measured  by AFM, with (c) showing the difference between the experimental and predicted heights. (d) The experimental peak shift, (e) the calculated peak shift using the experimentally determined strain gauge factors, and (f) the difference between the two.}
  \label{sample_20}
\end{figure}

\begin{figure}[h]
  \includegraphics[width=1\linewidth]{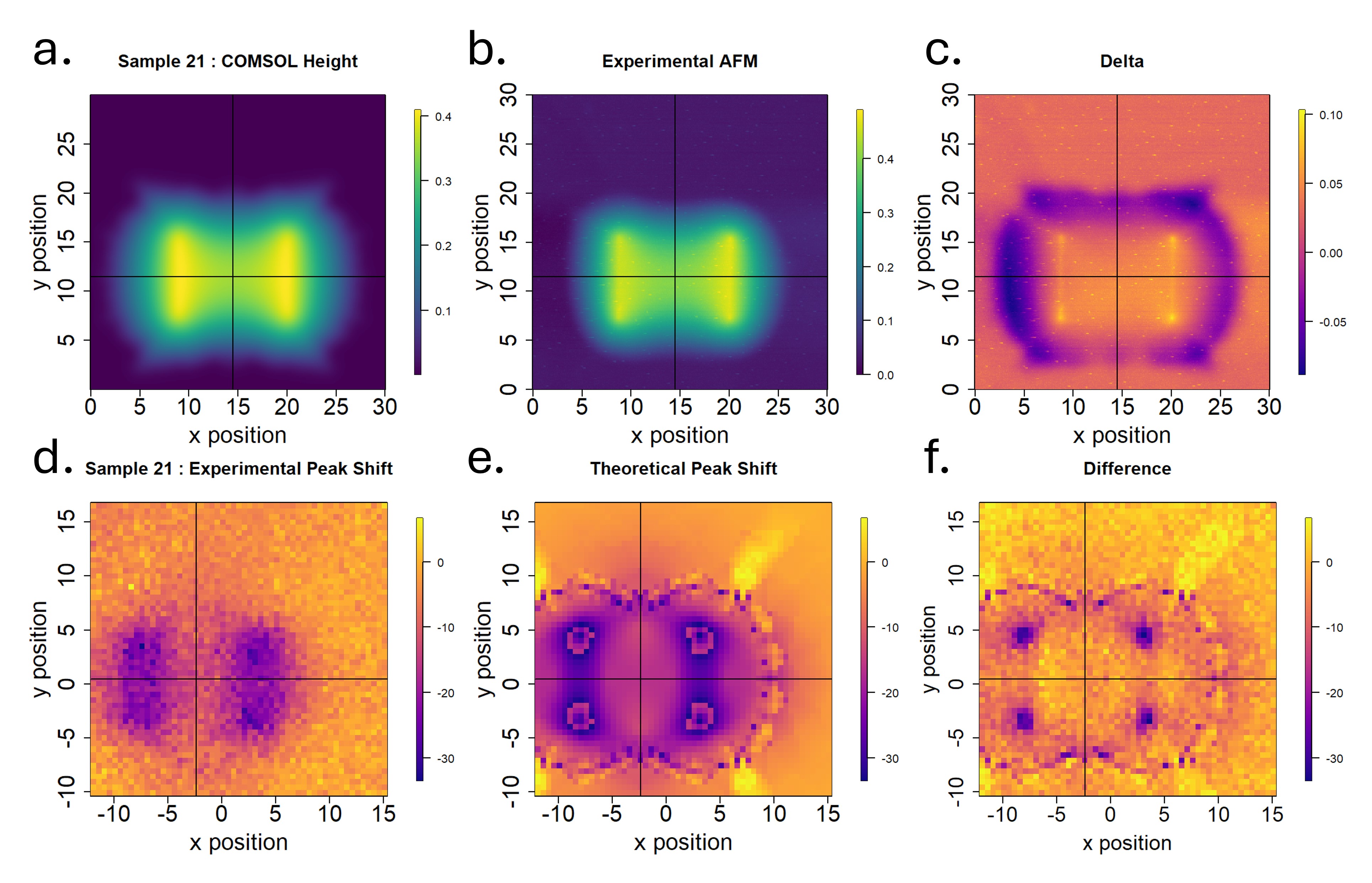}
  \caption{Sample 21. The height profile (a) simulated from COMSOL and (b) measured  by AFM, with (c) showing the difference between the experimental and predicted heights. (d) The experimental peak shift, (e) the calculated peak shift using the experimentally determined strain gauge factors, and (f) the difference between the two.}
  \label{sample_21}
\end{figure}
\begin{figure}[h]
  \includegraphics[width=1\linewidth]{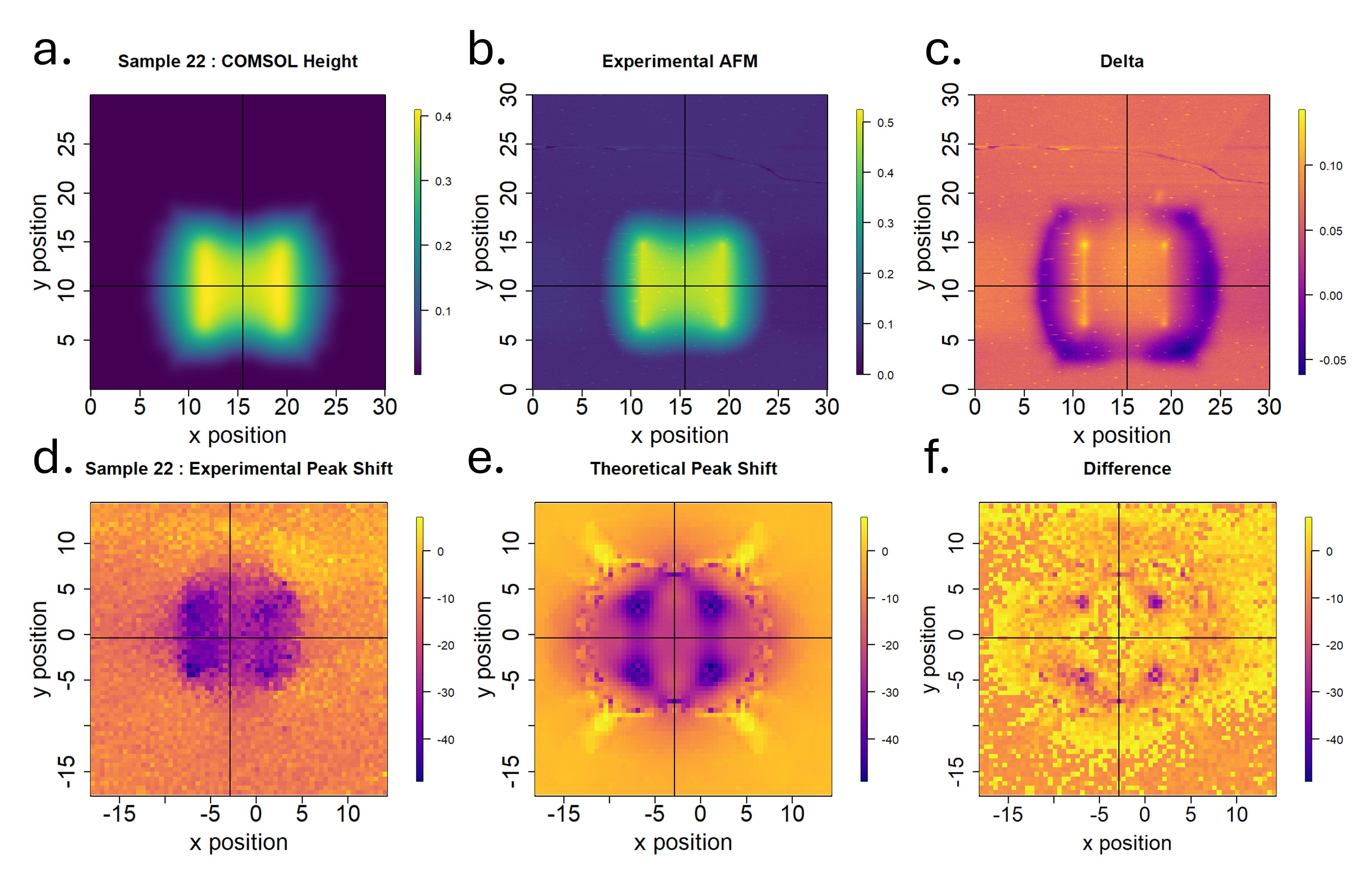}
  \caption{Sample 22. The height profile (a) simulated from COMSOL and (b) measured  by AFM, with (c) showing the difference between the experimental and predicted heights. (d) The experimental peak shift, (e) the calculated peak shift using the experimentally determined strain gauge factors, and (f) the difference between the two.}
  \label{sample_22}
\end{figure}

\bibliography {ref}